\documentclass[fleqn,usenatbib]{mnras}

\usepackage{newtxtext,newtxmath}

\usepackage[T1]{fontenc}

\usepackage{graphicx}	
\usepackage{amsmath}	

\usepackage[dvipsnames]{xcolor} 
\usepackage{hyperref} 
\usepackage{makecell} 
\usepackage[para]{threeparttable} 

\newcommand{\ie}{i.\,e.}

\newcommand{\teff}{$T_{{\rm eff}}$}
\newcommand{\kms}{\mbox{km\,s$^{-1}$}}
\newcommand{\ms}{\mbox{m s$^{-1}$}}
\newcommand{\gcm}{\mbox{g cm$^{-3}$}}

\newcommand{\vsini} {$v$\,sin\,$i$}

\newcommand{\gfeh} {\mbox{$[{\rm Fe}/{\rm H}]$}}

\newcommand{\logg} {\mbox{log\,{\it g}}}

\newcommand{\mplanet}{\mbox{$M_{\rm p}$}}
\newcommand{\rplanet}{\mbox{$R_{\rm p}$}}
\newcommand{\rhoplanet}{\mbox{$\rho_{\rm p}$}}

\newcommand{\mearth}{\mbox{M$_\oplus$}}
\newcommand{\rearth}{\mbox{R$_\oplus$}}
\newcommand{\rhoearth}{\mbox{$\rho_\oplus$}}

\newcommand{\msun}{\mbox{M$_\odot$}}
\newcommand{\rsun}{\mbox{R$_\odot$}}
\newcommand{\mstar}{\mbox{$M_\star$}}
\newcommand{\rstar}{\mbox{$R_\star$}}

\newcommand{\rhostar}{\mbox{$\rho_\star$}}
\newcommand{\rhosun}{\mbox{$\rho_\odot$}}

\newcommand{\logRHK}{\mbox{$\log {\rm R}^{\prime}_{\rm HK}$}}

\newcommand{\starname}{\mbox{TOI-561}}
\newcommand{\ticname}{\mbox{TIC 377064495}}

\newcommand{\tess}{\mbox{{\it TESS}}}
\newcommand{\cheops}{\mbox{{\it CHEOPS}}}

\title[The TOI-561 system as seen by CHEOPS, HARPS-N and TESS]{Investigating the architecture and internal structure of the TOI-561 system planets with CHEOPS, HARPS-N and TESS}

\author[G. Lacedelli et al.]{
G. Lacedelli$^{1,2}$\thanks{E-mail: gaia.lacedelli@phd.unipd.it},
T. G. Wilson$^{3}$, 
L. Malavolta$^{1,2}$, 
M. J. Hooton$^{4}$, 
A. Collier Cameron$^{3}$, 
Y. Alibert$^{4,5}$, 
\newauthor
A. Mortier$^{6,7}$, 
A. Bonfanti$^{8}$, 
R. D. Haywood$^{9}$, 
S. Hoyer$^{10}$
% ...
\newauthor
\emph{\normalsize Affiliations are listed at the end of the paper}}

\date{Accepted 2022 January 18. Received 2022 January 14; in original form 2021 November 25}

\pubyear{2022}

\begin{document}
\label{firstpage}
\pagerange{\pageref{firstpage}--\pageref{lastpage}}
\maketitle

\clearpage
\newpage

\begin{abstract}
We present a precise characterization of the TOI-561 planetary system obtained by combining previously published data with \tess\ and \cheops\ photometry, and a new set of $62$ HARPS-N radial velocities (RVs).
Our joint analysis confirms the presence of four transiting planets, namely TOI-561 b ($P = 0.45$~d, $R = 1.42$~\rearth, $M = 2.0$~\mearth), c ($P = 10.78$~d, $R = 2.91$~\rearth, $M = 5.4$~\mearth), d ($P = 25.7$~d, $R = 2.82$~\rearth, $M = 13.2$~\mearth) and e ($P = 77$~d, $R = 2.55$~\rearth, $M = 12.6$~\rearth). Moreover, we identify an additional, long-period signal ($>450$~d) in the RVs, which could be due to either an external planetary companion or to stellar magnetic activity.
The precise masses and radii obtained for the four planets allowed us to conduct interior structure and atmospheric escape modelling.
TOI-561 b is confirmed to be the lowest density ($\rho_{\rm b} = 3.8 \pm 0.5$~\gcm) ultra-short period (USP) planet known to date, and the low metallicity of the host star makes it consistent with the general bulk density-stellar metallicity trend. According to our interior structure modelling, planet b has basically no gas envelope, and it could host a certain amount of water.
In contrast, TOI-561 c, d, and e likely retained an H/He envelope, in addition to a possibly large water layer.
The inferred planetary compositions suggest different atmospheric evolutionary paths, with planets b and c having experienced significant gas loss, and planets d and e showing an atmospheric content consistent with the original one.
The uniqueness of the USP planet, the presence of the long-period planet TOI-561 e, and the complex architecture make this system an appealing target for follow-up studies. 
\end{abstract}



\section{Introduction}{\label{sec:intro}}
Since the announcement of the first exoplanet orbiting a Sun-like star \citep{mayor1995}, the growing number of discoveries in exoplanetary science have yielded a surprising variety of exoplanets and exoplanetary systems.
The field has benefited hugely from dedicated space-based missions, such as {\it CoRoT, Kepler, K2} \citep{baglin2006, Borucki2010, howell2014}, and recently {\it TESS} \citep{ricker2014}. With more than $170$ confirmed planets, and $\sim 4000$ planet candidates, the majority of which will likely turn out to be planets, {\it TESS} has increased the census of confirmed exoplanets to more than $4500$\footnote{From NASA Exoplanet Archive, \url{https://exoplanetarchive.ipac.caltech.edu/}.}.
Alongside the aforementioned missions, which are designed to discover a large number of exoplanets by searching for transit-like signatures around hundreds of thousands of stars, new characterization missions, with a specific focus on the detailed study of known exoplanets, are now starting to operate.
Among them, the {\it CHaracterising ExOPlanet Satellite} ({\it CHEOPS}, \citealt{benz2021}), launched on 18 December 2019, is a $30$~cm telescope which is collecting ultra-high precision photometry of known exoplanets, aiming at their precise characterization. {\it CHEOPS} met its precision requirements both on bright and faint stars, achieving a noise level of $\sim 15$~ppm per $6$~h intervals for $V \sim 9$~mag stars, and $75$~ppm per $3$~h for $V \sim 12$~mag stars \citep{benz2021}. The importance of such a high photometric precision is reflected in {\it CHEOPS}' first scientific results, which span a variety of different fields (\citealt{lendl2020, bonfanti2021, leleu2021, delrez2021, morris2021_55cnc, borsato2021, vangrootel2021, szabo2021, hooton2021, swayne2021, maxted2021,barros2022, wilson2022, deline2022}). 
As part of its main scientific goals, {\it CHEOPS} is refining the radii of known exoplanets to achieve the precision on the bulk density needed for internal structure and atmospheric evolution modelling. 
To fulfil this aim, radial velocity (RV) follow-ups using high-precision spectrographs are essential to provide the precise planetary masses that can be combined with radii measurements to determine accurate densities.
Among the exoplanets having both radius and mass measurements, the ones in well-characterised multiplanetary systems are of particular interest, since they allow for investigation of their formation and evolution processes through comparative planetology, e.g. by comparing their individual inner bulk compositions (e.g. \citealt{Guenther2017, prieto2018}), by studying their mutual inclinations and eccentricities (e.g. \citealt{Fabrycky_2014, vanEylen2019, mills2019}), and by investigating the correlations between their relative sizes, masses and orbital separations (e.g. \citealt{Lissauer_2011, ciardi2013, millholland2017, weiss2018, jiang2020, adams2020}).

Within this context, TOI-561, announced simultaneously by \citet{lacedelli2021} and \citet{weiss2021} (L21 and W21 hereafter, respectively), is a particularly interesting system, both from the stellar (Section~\ref{sec:star}) and planetary (Section~\ref{sec:pl_system}) perspective. 
The low stellar metallicity, the presence of an ultra-short period (USP) planet, where USP planets are meant here as planets with periods shorter than one day and radii smaller than $2$~\rearth, and the complexity of its planetary configuration make TOI-561 an appealing target for in-depth investigations. 
In this study, we combine literature data with new {\it TESS} observations (Section~\ref{sec:tess_observations}), {\it CHEOPS} photometry (Section~\ref{sec:cheops_observations}), and HARPS-N RVs (Section~\ref{sec:harpn_observations}) to shed light on the planetary architecture and infer the internal structure of the transiting planets.
After assessing the planetary configuration using {\it CHEOPS} observations (Section~\ref{sec:pl_d}) and performing a thorough analysis of the global RV data set (Section~\ref{sec:additional_pl}), we jointly modelled the photometric and spectroscopic data to obtain the planetary parameters (Section~\ref{sec:global_analysis}). 
We used our derived stellar and planetary properties to model the internal structures of the transiting planets (Section~\ref{sec:internal_model}) and their atmospheric evolution (Section~\ref{sec:atmo_model}), before discussing our results and presenting our conclusions (Section~\ref{sec:conclusion}).

\section{The TOI-561 system}{\label{sec:toi561}}
\subsection{The host star}{\label{sec:star}}
TOI-561 is an old, metal-poor, thick disk star (L21, W21), slightly smaller and cooler than the Sun, located $\sim 84$~pc away from the Solar System. We report the main astrophysical properties of the star in Table~\ref{table:star_params}.

We adopted the spectroscopic parameters and stellar abundances from L21 (Table~\ref{table:star_params}), which were derived exploiting the high SNR, high-resolution HARPS-N co-added spectrum (L21, \S~3.1) through an accurate analysis using three independent methods, namely the ARES+MOOG equivalent width method \citep{Sousa-14, Mortier2014}, the Stellar Parameter Classification \citep{Buchhave2012, Buchhave2014} and the CCFpams method \citep{Malavolta2017b}.

Taking advantage of the updated parameters coming from the {\it Gaia} EDR3 release \citep{GaiaCollab2021}, we then used the L21 spectral parameters as priors on spectral energy distribution selection to infer the stellar radius (\rstar) of TOI-561 using the infrared flux method (IRFM, \citealt{blackwell1977}). 
The IRFM compares optical and infrared broadband fluxes and synthetic photometry of stellar atmospheric models, and uses known relationships between stellar angular diameter, \teff\ and parallax to derive \rstar, in a MCMC fashion as detailed in \citet{Schanche2020}. 
For this study, we retrieved from the most recent data releases the {\it Gaia G}, {\it G$_{\rm BP}$}, {\it G$_{\rm RP}$} \citep{GaiaCollab2021}, 2MASS $J$, $H$, $K$ \citep{Skrutskie2006}, and {\it WISE} $W1$, $W2$ \citep{Wright2010} broadband photometric magnitudes, and we used the stellar atmospheric models from the \texttt{ATLAS} Catalogues \citep{Castelli2003} and the {\it Gaia} EDR3 parallax with the offset of \citet{Lindegren2021} applied, to obtain \rstar~$= 0.843 \pm 0.005$~\rsun.

We combined two different sets of stellar evolutionary tracks and isochrones, PARSEC\footnote{http://stev.oapd.inaf.it/cgi-bin/cmd} \citep[\textit{PA}dova \& T\textit{R}ieste \textit{S}tellar \textit{E}volutionary \textit{C}ode, v1.2S;][]{Marigo_2017} and CLES \citep[Code Liègeois d’Évolution Stellaire,][]{Scuflaire2008}, to derive the stellar mass ($M_{\star}$) and age ($t_{\star}$) of TOI-561.
As the star is significantly alpha-enhanced, we avoided using $[$Fe/H$]$ as a proxy for the stellar metallicity; instead, we inserted both $[$Fe/H$]$ and [$\alpha$/Fe] in relation (3) provided by \citet{yi01}, obtaining an overall scaling of metal abundances $\mathrm{[M/H]}=-0.23\pm0.06$.
Besides $[$M/H$]$, the main input parameters for computing $M_{\star}$ and $t_{\star}$ were $T_{\mathrm{eff}}$ and $R_{\star}$. 
In addition, we used as inputs \logRHK\ and the upper limit on \vsini\ from L21, and the yttrium over magnesium abundance $\mathrm{[Y/Mg]}=-0.22\pm0.07$, as computed from $[$Mg/H$]$ and $[$Y/H$]$ reported by W21. 
These indices improve the model convergence by discarding unlikely young isochrones, as broadly discussed in Section 2.2.3 of \citet{bonfanti20}, and references therein.
The PARSEC results were obtained using the isochrone placement algorithm of \citet{bonfanti2015, bonfanti2016}, which retrieves the best-fit parameters by interpolating within a pre-computed grid of models, while the CLES algorithm models directly the star through a Levenberg-Marquardt minimisation \citep{salmon2021}. 
The final adopted values ($M_{\star}=0.806 \pm 0.036$~\msun, $t_{\star}=11.0_{-3.5}^{+2.8}$~Gyr) are a combination of the outputs from both sets of models, as described in detail in \citet{bonfanti2021}. 
The derived mass and radius, listed in Table~\ref{table:star_params}, are consistent within $1\sigma$ with the values reported in L21 ($\rstar = 0.849 \pm 0.007$~\rsun, $\mstar = 0.785 \pm 0.018$~\msun).

\begin{table}
\caption{Stellar properties of \starname.}
\label{table:star_params}
\begin{threeparttable}[t]
\centering
\begin{tabular}{lll}
\hline\hline
\multicolumn{3}{c}{TOI-561}\\
\hline
TIC & \multicolumn{2}{l}{377064495} \\
{\it Gaia} EDR3 & \multicolumn{2}{l}{3850421005290172416}\\
2MASS & \multicolumn{2}{l}{J09524454+0612589} \\[1ex]
\hline
Parameter & Value & Source \\
\hline
RA  (J2016; hh:mm:ss.ss) &  09:52:44.43 & A\\
Dec (J2016; dd:mm:ss.ss) & $+$06:12:57.94 & A\\
$\mu_{\alpha}$ (mas yr$^{-1}$) & $-108.504 \pm 0.022$ & A\\
$\mu_{\delta}$ (mas yr$^{-1}$) & $-61.279 \pm 0.019$ & A\\
$\gamma$ (\kms) & $79.54 \pm 0.56$ & A \\
Parallax (mas) & $11.8342 \pm 0.0208 $& A\\
Distance (pc) & $ 84.25 \pm 0.12 $& B \\
\hline
{\it TESS} (mag) & $9.527 \pm 0.006$& C\\
{\it G} (mag) & $10.0181 \pm 0.0028$ & A\\
{\it G$_{\rm BP}$} (mag) & $10.3945 \pm 0.0028$ & A\\
{\it G$_{\rm RP}$} (mag) & $9.4692 \pm 0.0038$ & A\\
{\it V } (mag) & $10.252 \pm 0.006$ & C\\
{\it B }(mag) & $10.965 \pm	0.082$ & C\\
{\it J }(mag) & $8.879  \pm	0.020$ & D\\
{\it H }(mag) & $8.504  \pm	0.055$ & D\\
{\it K }(mag) & $8.394  \pm	0.019$ & D\\
{\it W1} (mag) & $8.337  \pm	0.023$ & E\\
{\it W2} (mag) & $8.396  \pm	0.020$ & E\\[1ex]
\hline
\teff\ (K) & $ 5372 \pm 70 $ & F \\ 
\logg\  (cgs)  & $ 4.50 \pm 0.12$ & F \\ 
\gfeh\  (dex) & $ -0.40 \pm 0.05$ & F \\
$[$Mg/H$]$ (dex) & $-0.17 \pm 0.05$ & F \\
$[$Si/H$]$ (dex) & $-0.22 \pm 0.05$ & F \\
$[$Ti/H$]$ (dex) & $-0.12 \pm 0.03$ & F \\
$[\alpha$/Fe$]$ (dex) & $0.23 \pm 0.04$& F\\
$[$M/H$]$ (dex) & $-0.23\pm0.06$ & G \\
$[$Y/Mg$]$ (dex) & $-0.22\pm0.07$ & G$^a$\\
\logRHK & $-5.003 \pm 0.012$ & F \\
\vsini\ (\kms) & $< 2$ & F\\
\rstar\ (\rsun) & $ 0.843 \pm 0.005 $ & G, IRFM \\
\mstar\ (\msun) & $ 0.806 \pm 0.036 $ & G, isochrones \\
$t_{\star}$ (Gyr)  & $11.0_{-3.5}^{+2.8}$ & G, isochrones \\
\rhostar\ (\rhosun)& $ 1.34 \pm 0.06$ & G, from \rstar\ and \mstar \\
\rhostar\ (\gcm) & $1.89 \pm 0.09$ & G, from \rstar\ and \mstar \\
$L_{\star}$ ($L_{\odot}$) & $0.533 \pm 0.029$& G, from \rstar\ and \teff\\ 
Spectral type & G9V & F\\
\hline\hline
\end{tabular}
\begin{tablenotes}
A) {\it Gaia} EDR3 \citep{GaiaCollab2021}. B) \citet{bailer_jones2021}.
C) {\it TESS} Input Catalogue Version 8 (TICv8, \citealt{Stassun2018}).
D) Two Micron All Sky Survey (2MASS, \citealt{Cutri2003}).
E) {\it Wide-field Infrared Survey Explorer} \citep[{\it WISE};][]{Wright2010}.
F) L21. G) This work. \\
$^a$ Based on W21 abundances.
\end{tablenotes}
\end{threeparttable}
\end{table}

\subsection{The planetary system}{\label{sec:pl_system}}
The discovery of a multiplanetary system orbiting TOI-561 was announced simultaneously by L21 and W21 in two independent papers.
The main planetary parameters from both studies are reported in Table~\ref{table:literature_pl}.

The two papers presented different RV data sets, collected with HARPS-N and HIRES respectively, to confirm the planetary nature of three candidates identified by {\it TESS} in sector $8$, the only available sector at the time of the publications.
The {\it TESS}-identified signals had periods of $ \sim 0.45$, $\sim 10.8$, and $\sim 16$ days. 
The two inner candidates were confirmed by both L21 and W21, with the names of TOI-561 b (an USP super-Earth, with period $P_{\rm b} \sim 0.4465$~d, and radius $R_{\rm b} \sim 1.4$~\rearth), and TOI-561 c (a warm mini-Neptune, with $P_{\rm c} \sim 10.779$~d, and $R_{\rm c} \sim 2.9$~\rearth). 
However, two different interpretations for the third {\it TESS} signal were proposed by the authors. 
In the scenario presented in L21, the two transits related to the third {\it TESS} signal were interpreted as single transits of two distinct planets, TOI-561 d ($P_{\rm d} \sim 25.6$~d, $R_{\rm d} \sim 2.5$~\rearth), and TOI-561 e ($P_{\rm e} \sim 77$~d, $R_{\rm e} \sim 2.7$~\rearth). 
The periods of these two planets were inferred from the RV analysis, which played an essential role in determining the final planetary architecture.
In fact, the ephemeris match between the RV and photometric fits (See Fig.~5 of L21) and the non-detection of the $16$~d signal in the HARPS-N data set, combined with the different durations of the two {\it TESS} transits and results from the long-term stability analysis led the authors to converge on a 4-planet configuration, presenting robust mass and radius detection for all the four planets in the system (L21, Table 5).
In contrast, W21 proposed the presence of a single planet at the period suggested by {\it TESS} (TOI-561 f, $P_{\rm f} \sim 16.29$~d, $R_{\rm f} \sim 2.3$~\rearth), based on the analysis of the two available transits.
W21 pointed out that the $8.1$~d alias of planet f's orbital period is also consistent with the \tess\ data, with the even transit falling into the \tess\ download gap, even though in this case the transit duration would be too long compared to
what is expected for a 8~d period planet (\S 4.9, W21).
However, the authors could not obtain an accurate mass determination for this planet, with the $60$ HIRES RVs being consistent with a non-detection (\S~7.2, W21). 
An additional discrepancy between the two studies is the mass of the USP planet, differing by almost a factor two. 
According to the W21 analysis, TOI-561 b has a mass of $ 3.2 \pm 0.8$~\mearth, making it consistent with a rocky composition and placing it among the population of typical small ($< 2$~\rearth), extremely irradiated USP planets \citep{Sanchis_Ojeda_2015, Dai_2021}.
Instead, assuming the low mass ($M_{\rm b} = 1.59 \pm 0.36$~\mearth) inferred from L21 analysis, TOI-561 b is not consistent with a pure rocky composition, and it is the lowest density USP super-Earth known to date, calling for a more complex interpretation (e.g. lighter core composition, deep water reservoirs, presence of a high-metallicity, volatile materials or water steam envelope, etc.).

The complexity of this system and the differences between the two studies demanded further investigations. 
We therefore decided to collect additional, precise photometric and RV data (Section~\ref{sec:observations}) to shed light on the planetary configuration and on the internal composition of the TOI-561 planets. 

\begin{table}
\caption{Literature parameters of the proposed planets orbiting \starname.}
\label{table:literature_pl}
\begin{threeparttable}[t]
\centering
\begin{tabular}{lll}
\hline\hline
TOI-561 b & \citet{lacedelli2021} & \citet{weiss2021} \\
\hline
$P$ (d) & $	0.446578 \pm 0.000017$ & $0.446573^{+0.000032}_{-0.000021}$\\
$T_0$ (TBJD) & $1517.498 \pm 0.001$ & $1517.4973 \pm 0.0018$ \\
$R_{\rm p}$ (\rearth) & $1.423 \pm 0.066$ & $1.45 \pm 0.11$\\
$K$ (\ms) & $1.56 \pm 0.35$ & $3.1 \pm 0.8$\\
$M_{\rm p}$ (\mearth) & $1.59 \pm 0.36$ & $3.2 \pm 0.8$ \\[1ex]
\hline
TOI-561 c &  &\\
\hline
$P$ (d) & $10.779 \pm 0.004$ & $10.77892 \pm 0.00015$ \\
$T_0$ (TBJD) & $1527.060 \pm 0.004$ & $1527.05825 \pm 0.00053$ \\
$R_{\rm p}$ (\rearth) & $2.878 \pm 0.096$ & $2.90 \pm 0.13$\\
$K$ (\ms) & $1.84 \pm 0.33$ & $2.4 \pm 0.8$\\
$M_{\rm p}$ (\mearth) & $5.40 \pm 0.98$& $7.0 \pm 2.3$\\[1ex]
\hline
TOI-561 d & & \\
\hline
$P$ (d) & $25.62 \pm 0.04$ & - \\
$T_0$ (TBJD) & $1521.882 \pm 0.004$ & - \\
$R_{\rm p}$ (\rearth) & $2.53 \pm 0.13$& - \\
$K$ (\ms) & $3.06 \pm 0.33$ & -\\
$M_{\rm p}$ (\mearth) & $11.95 \pm 1.28$ & - \\[1ex]
\hline
TOI-561 e & & \\
\hline
$P$ (d) & $77.23 \pm 0.39$& -\\
$T_0$ (TBJD) & $8538.181 \pm 0.004$ & -\\
$R_{\rm p}$ (\rearth) & $2.67 \pm 0.11$ & -\\
$K$ (\ms) & $2.84 \pm 0.41$ & -\\
$M_{\rm p}$ (\mearth) & $16.0 \pm 2.3$ & -\\[1ex]
\hline
TOI-561 f$ \, ^a$ & & \\
\hline
$P$ (d) & - & $16.287 \pm 0.005$\\
$T_0$ (TBJD) & - & $1521.8828 \pm 0.0035$\\
$R_{\rm p}$ (\rearth) & - & $2.32 \pm 0.16$\\
$K$ (\ms) & - & $0.9 \pm 0.6$\\
$M_{\rm p}$ (\mearth) & - & $3.0^{+2.4}_{-1.9}$\\[1ex]
\hline
\hline
$\mathrm{\mstar\ (\msun)}$ & $0.785 \pm 0.018$ & $0.805 \pm 0.030$\\
\hline
\end{tabular}
\begin{tablenotes}
$^a$ Referred as TOI-561 d in W21.
\end{tablenotes}
\end{threeparttable}
\end{table}

\section{Observations}{\label{sec:observations}}
\subsection{{\it TESS} photometry}\label{sec:tess_observations}
\begin{figure*}
\centering
  \includegraphics[width=\linewidth]{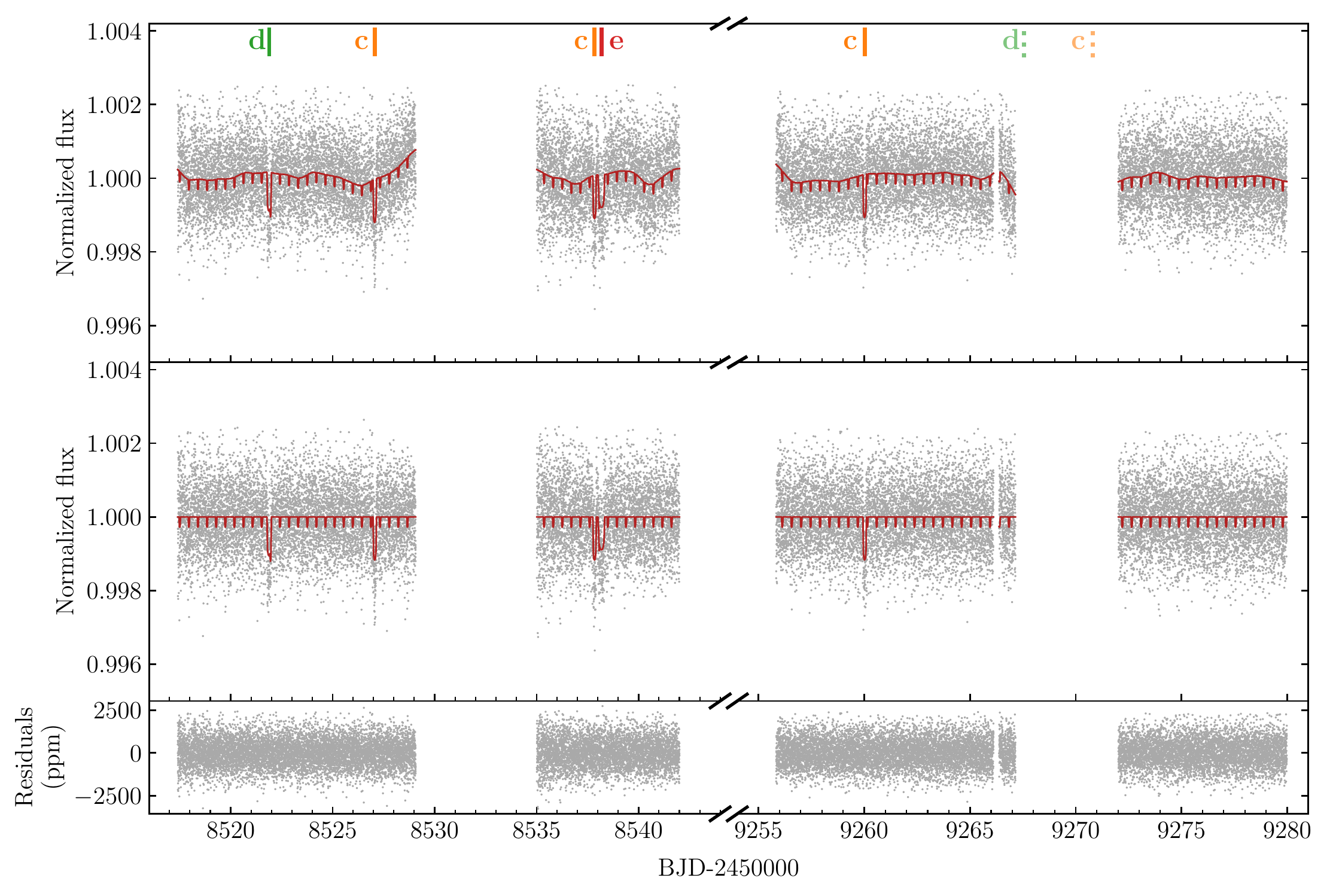}
  \caption{{\it TESS} sector $8$ ({\it left}) and $35$ ({\it right}) PDCSAP light curves of TOI-561. In the top panel, the dark red solid line shows the best-fitting transit and Matérn-3/2 kernel Gaussian Process (GP) model, as detailed in Section~\ref{sec:global_analysis}. The central panel shows the flattened light curve after the removal of the GP component, with the best-fitting transit model superimposed (dark red solid line). The transits of planets c, d and e are labelled and highlighted with orange, green and red vertical lines, respectively. The expected locations of the transits of planets c and d occurring during the data gaps of sector 35 are marked with pale, dashed orange and green lines, respectively. Planet e is not expected to transit in sector 35. The transits of the USP planet are too shallow to be individually visible, and are not indicated. Light curve residuals are shown in the bottom panel.
  }
    \label{fig:tess_lc}
\end{figure*}

During its two-year primary mission \citep{ricker2014}, {\it TESS} observed \starname\  in two-minute cadence mode between 2 February and 27 February 2019 (sector $8$).
After entering its extended mission, {\it TESS} re-observed the star in two-minute cadence mode during sector $35$, between 9 February and 6 March 2021. 
At the beginning of the second orbit, the spacecraft dropped out of
Fine Pointing mode for $3.44$ days, entering Coarse Pointing mode\footnote{See {\it TESS} Data Release Notes: Sector $35$, DR51 (\url{https://archive.stsci.edu/tess/tess_drn.html}).}. Data collected during Coarse Pointing mode were flagged and removed from the Pre-search Data Conditioning Simple Aperture Photometry \citep[PDCSAP, ][]{smith2012, Stumpe2012, Stumpe2014} light curves, leading to a total of $19.86$ days of science data.
The photometric observations of \starname\
were reduced by the Science Processing Operations Center (SPOC) 
pipeline and searched for evidence of transiting planets \citep{jenkins2016,jenkins2020}.
For our photometric analysis, 
we used the light curves based on the PDCSAP, downloading the two-minute cadence data from the Mikulski Archive for Space Telescopes (MAST)\footnote{
\url{https://mast.stsci.edu/portal/Mashup/Clients/Mast/Portal.html}},
and removing all the observations
encoded as \emph{NaN} or flagged as bad-quality (\texttt{DQUALITY>0}) points by the SPOC pipeline\footnote{ \url{https://archive.stsci.edu/missions/tess/doc/EXP-TESS-ARC-ICD-TM-0014-Rev-F.pdf}}. We performed outlier rejection by doing a cut at $3 \sigma$ for positive outliers and $5 \sigma$ (\ie\ larger than the deepest transit) for negative outliers. 
The resulting {\it TESS} light curves of sectors 8 and 35 are shown in Figure~\ref{fig:tess_lc}, and Table~\ref{table:tess_log} summarizes the total number of transits observed by \tess\ for each planet.

To refine the ephemeris of planet d in time for the scheduling of the {\it CHEOPS} observations (Section~\ref{sec:pl_d}), we also extracted the $10$-minute
cadence light curve of sector $35$ using the quick-look \tess\ Full Frame Images (FFIs) calibrated using the \tess\ Image CAlibrator\footnote{\url{https://archive.stsci.edu/hlsp/tica}} package (\texttt{tica}, \citealt{fausnaugh2021}).

\begin{table}
  \caption{ Number of TOI-561 transits observed by {\it TESS}.}
\label{table:tess_log}      
\centering                                      
\begin{tabular}{l c c c c}          
\hline\hline                        
 & TOI-561 b & TOI-561 c & TOI-561 d & TOI-561 e \\
\hline                                 
Sector 8  & 41 & 2 & 1 & 1 \\
Sector 35 & 43 & 1 & - & - \\
\hline
\end{tabular}
\end{table}

\subsection{CHEOPS photometry}\label{sec:cheops_observations}
To confirm the planetary architecture and improve the planetary parameters, we obtained three visits of \starname\ with {\it CHEOPS}, the ESA small class mission dedicated to the characterization of known exoplanets \citep{benz2021}. 
The observations, collected within the Guaranteed Time Observing (GTO) programme, were carried out between 23 January and 15 April 2021, for a total of $73.85$ hours on target. 
During the three visits, we observed a total of eight transits of TOI-561 b, two transits of TOI-561 c, and one transit of TOI-561 d.
The three {\it CHEOPS} light curves have an observing efficiency, i.e. the actual time spent observing the target with respect to the total visit duration, of 64\%, 75\%, and 61\%, respectively.
The observing efficiency is linked to data gaps, which are intrinsically present in all {\it CHEOPS} light curves (see e.g. \citealt{delrez2021}, \citealt{bonfanti2021}, \citealt{leleu2021}), and are related to \cheops's low-Earth orbit. In fact, during (1) South Atlantic Anomaly (SAA) crossing, (2) target occultation by the Earth, and (3) too high stray light contamination, no data are downlinked. This results in data gaps, whose number and extension depend on the target sky position \citep{benz2021}.
For all the visits, we adopted an exposure time of $60$~s.
The summary log of the {\it CHEOPS} observations is reported in Table~\ref{table:cheops_log}.

\begin{table*}
  \caption{Log of TOI-561 {\it CHEOPS} observations.}
\label{table:cheops_log}      
\centering                                      
\begin{tabular}{c c c c c c c c}          
\hline\hline                        
Visit & File key & Starting date & Duration & Data points & Efficiency & Exposure time & Planets \\
(\#) & & (UTC) & (h) & (\#) & (\%) & (s) & \\
\hline                                 
1 & CH\_PR100031\_TG037001\_V0200 & 2021-01-23T15:29:07 & 15.67 & 604  & 64 & 60 & b,c \\
2 & CH\_PR100008\_TG000811\_V0200  & 2021-03-29T10:19:08 & 4.42  & 207  & 75 & 60 & b,c \\
3 & CH\_PR100031\_TG039301\_V0200 & 2021-04-12T23:52:28 & 53.76 & 1978 & 61 & 60 & b,d \\\hline
\end{tabular}
\end{table*}

Data were reduced using the latest version of the {\it CHEOPS} automatic Data Reduction Pipeline (DRP v13; \citealt{hoyer2020}), which performs aperture photometry of the target after calibrating the raw images (event flagging, bias, gain, non-linearity, dark current, and flat field) and correcting them for instrumental and environmental effects (smearing trails, cosmic rays, de-pointing, stray light, and background).
The target flux is obtained for a set of three fixed-radius apertures, namely $R = 22.5$~arcsec (RINF), $25.0$~arcsec (DEFAULT), $30.0$~arcsec (RSUP), plus an additional one specifically computed to optimize the radius based on the instrumental noise and contamination level of each target (OPTIMAL).
Moreover, the DRP estimates the contamination in the photometric aperture due to nearby targets using the sources listed in the {\it Gaia} DR2 catalog \citep{GaiaColl2018} to simulate the {\it CHEOPS} Field-of-View (FoV) of the target, as described in detail in \citet{hoyer2020}. 
No strong contaminants are present in the TOI-561 FoV, and the main contribution to the contamination is due to the smearing trails of a $G = 10.20$~mag star at a projected sky distance of $\sim 117.9$~arcsec, which rotates around the target inside the CCD window because of the {\it CHEOPS} field rotation \citep{benz2021}.
During the third visit three telegraphic pixels (pixels with a non-stable and abnormal behaviour during the visit) appeared within the {\it CHEOPS} aperture, one of them inside the {\it CHEOPS} PSF (Figure~\ref{fig:cheops_fov}). A careful treatment, described in detail in Appendix~\ref{sec:app_tel_pix}, was applied to correct for their effect.
In the subsequent analysis we adopted for all the visits the RINF photometry (see Figure~\ref{fig:cheops_raw} in Appendix~\ref{sec:app_tel_pix}), which minimized the light curve root mean square (RMS) dispersion, and we removed the outliers by applying a $4\sigma$ clipping.

\begin{figure}
\centering
  \includegraphics[width=\linewidth]{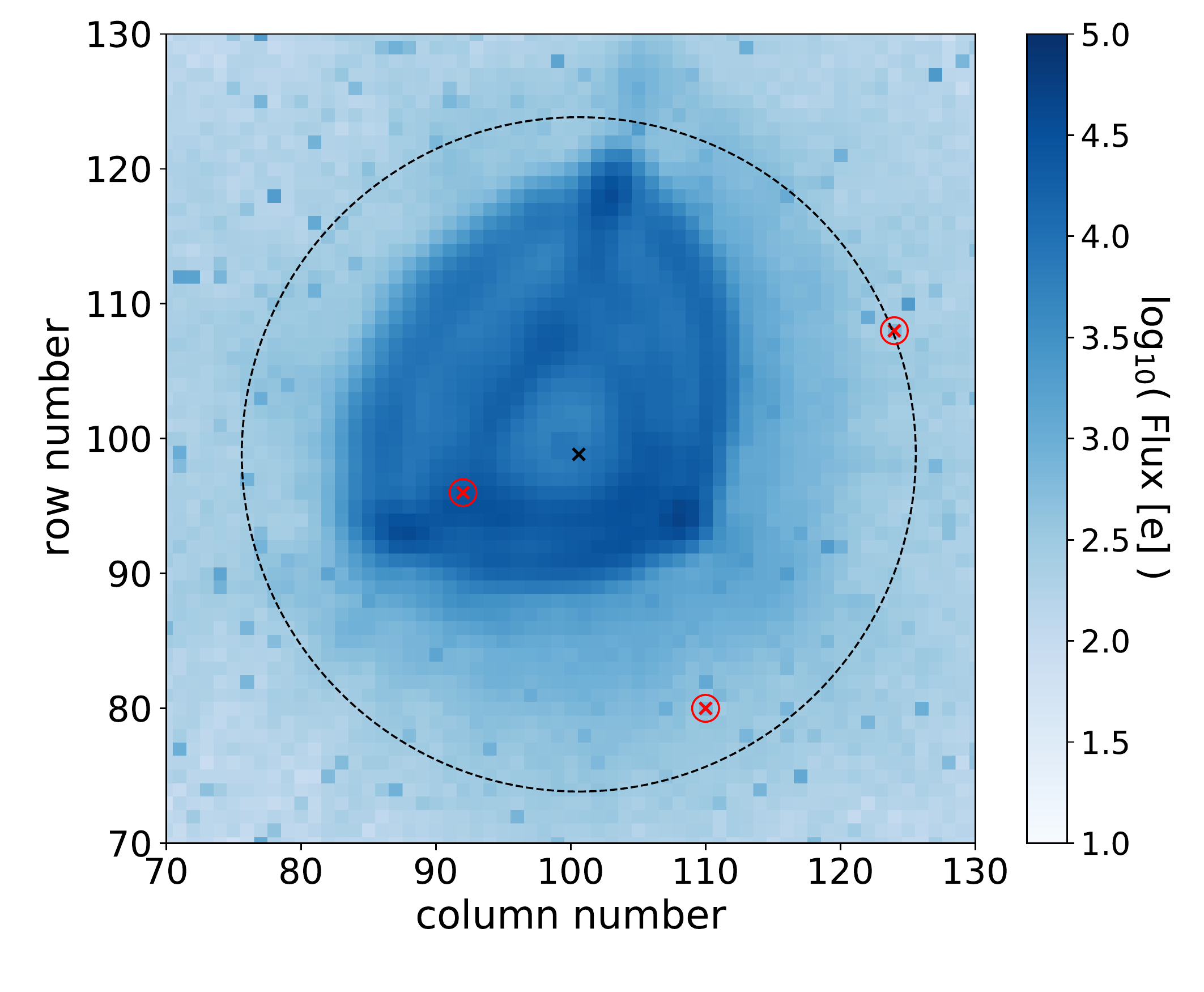}
  \caption{Extraction of $60\times 60$~arcsec of the CHEOPS FoV during the third visit centered on TOI-561. The dashed black circle represents the RINF photometric aperture surrounding the {\it CHEOPS} PSF, whose centroid is marked by the black cross. The positions of the three identified telegraphic pixels, including the one located within the {\it CHEOPS} PSF (see Appendix~\ref{sec:app_tel_pix}), are highlighted by the red, circled crosses. 
    }
    \label{fig:cheops_fov}
\end{figure}

\begin{figure*}
\centering
  \includegraphics[width=\linewidth]{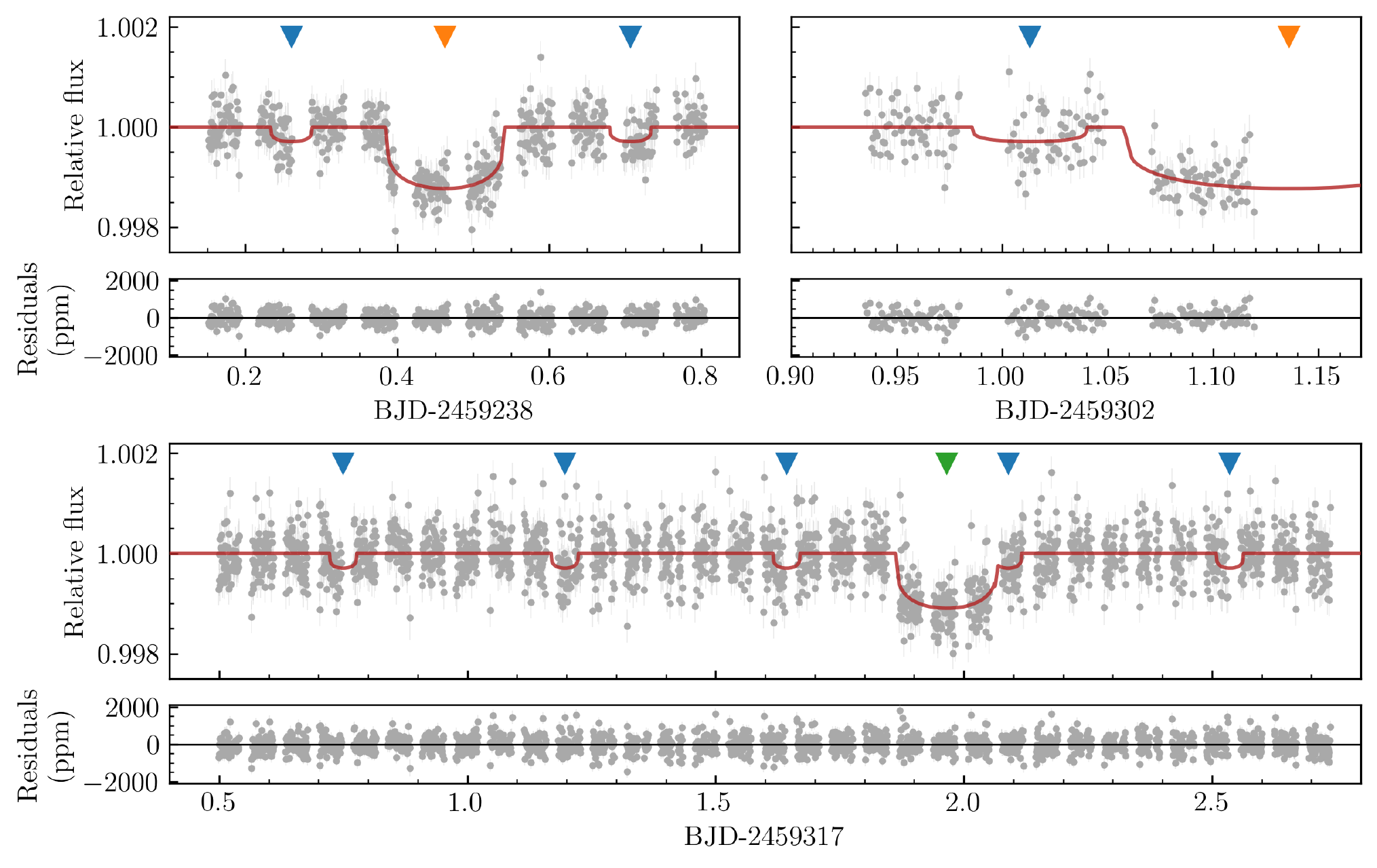}
  \caption{{\it CHEOPS} detrended light curves of TOI-561. Visits $1$, $2$ and $3$ are shown in the top left, top right, and bottom panel, respectively. The best-fitting model is over-plotted as a red solid line, and residuals are shown for each visit. The transits of planets b, c, and d are highlighted with blue, orange, and green triangles, respectively.
    }
    \label{fig:cheops_lc}
\end{figure*}

Finally, a variety of non-astrophysical sources, such as varying background, nearby contaminants or others, can produce short-term photometric trends in the {\it CHEOPS} light curves on the timescale of one orbit, due to the rotation of the {\it CHEOPS} FoV around the target and due to the nature of the spacecraft orbit. 
To correct for these effects, we detrended the light curves using the basis vectors provided by the DRP, as detailed in Section~\ref{sec:global_analysis}. The resulting detrended light curves are shown in Figure~\ref{fig:cheops_lc}.

\subsection{HARPS-N spectroscopy}\label{sec:harpn_observations}
In addition to the $82$ RVs published in L21, we collected $62$ high-resolution spectra using HARPS-N at the Telescopio Nazionale Galileo (TNG), in La Palma \citep{Cosentino2012, Cosentino2014}.
These were used to refine the planetary masses and confirm the system configuration.
The new observations were collected between 15 November 2020 and 1 June 2021. Following the same strategy of the previous season (L21), in addition to $30$ single observations, we collected six points per night on 8 and 10 February 2021, and two points per night on ten additional nights, specifically targeting the USP planet.
The exposure time for all the observations was set to $1800$~s, resulting in a signal-to-noise ratio (SNR) at $550$~nm of $83 \pm 20$ (median $\pm$ standard deviation) and a radial velocity measurement uncertainty of $1.0 \pm 0.4$~\ms.
All the observations were gathered with the second HARPS-N fibre illuminated by the Fabry–Perot calibration lamp
to correct for the instrumental RV drift.

We reduced the global HARPS-N data set ($144$ RVs in total) using the new version of the HARPS-N Data Reduction Software based on the ESPRESSO pipeline (DRS, version 2.3.1; see \citealt{dumusque2021} for more details). We used a G2 flux template to correct for variations in the flux distribution as a function of wavelength, and a G2 binary mask to compute the cross-correlation function (CCF, \citealt{Baranne1996, Pepe2002}).
We report the RVs and the associated activity indices (see Section~\ref{sec:additional_pl}) with their $1\sigma$ uncertainties in Table~\ref{table:harpn}. 
As in L21, we removed from the first season data set five RVs with associated errors $> 2.5$~\ms\ from spectra with SNR $< 35$ (see Appendix B in L21). All the RV uncertainties of the second season data set were below $2.5$ \ms, so no points were removed.

\begin{table*}
  \caption{HARPS-N radial velocity and activity indices measurements.}
\label{table:harpn}      
\centering                                      
\begin{tabular}{c c c c c c c c c c c c c}          
\hline\hline                        
  BJD$_{\rm TDB}$ & RV & $\sigma_{\rm RV}$ &  FWHM  & $\sigma_{\rm FWHM}$ & BIS & $\sigma_{\rm BIS}$& Contrast & $\sigma_{\rm contr}$ & S-index & $\sigma_{\text S}$ & H$\alpha$ & $\sigma_{\text H\alpha}$\\
 $($d$)$  & (\ms ) & (\ms ) & (\kms ) & (\kms ) & (\ms ) & (\ms ) & & & & & (dex) & (dex)\\
\hline                                 
2458804.70780  &  79695.97  &  1.13  & 6.415 & 0.002 & -86.82  &  2.26  & 59.879 & 0.021 & 0.1643 & 0.0005 & 0.2101 & 0.0002\\
2458805.77552  &  79699.66  &  0.85  & 6.419 & 0.002 & -85.13  &  1.71  & 59.810 & 0.016 & 0.1702 & 0.0003 & 0.2124 & 0.0001\\
2458806.76769  &  79697.50  &  0.91  & 6.415 & 0.002 & -83.66  &  1.82  & 59.861 & 0.017 & 0.1689 & 0.0003 & 0.2082& 0.0001\\
... & ... & ... & ... & ... & ... & ... & ... & ... & ... & ... & ... & ...\\
\hline
\multicolumn{9}{l}{This table is available in its entirety in machine-readable form.}
\end{tabular}
\end{table*}

\subsection{HIRES spectroscopy}\label{sec:hires_observations}
We included in our analysis $60$ high-resolution spectra collected with the W.M. Keck Observatory HIRES instrument on Mauna Kea, Hawaii between May 2019 and October 2020. The data set was published in W21, and we refer to that paper for details regarding the observing and data reduction procedures. The HIRES data set has an RMS of $5$~\ms, and a median individual RV uncertainty of $1.4$~\ms\ (W21).

\section{Probing the system architecture}{\label{sec:prob_config}}
\subsection{CHEOPS confirmation of TOI-561 d}
\begin{figure}
\centering
  \includegraphics[width=\linewidth]{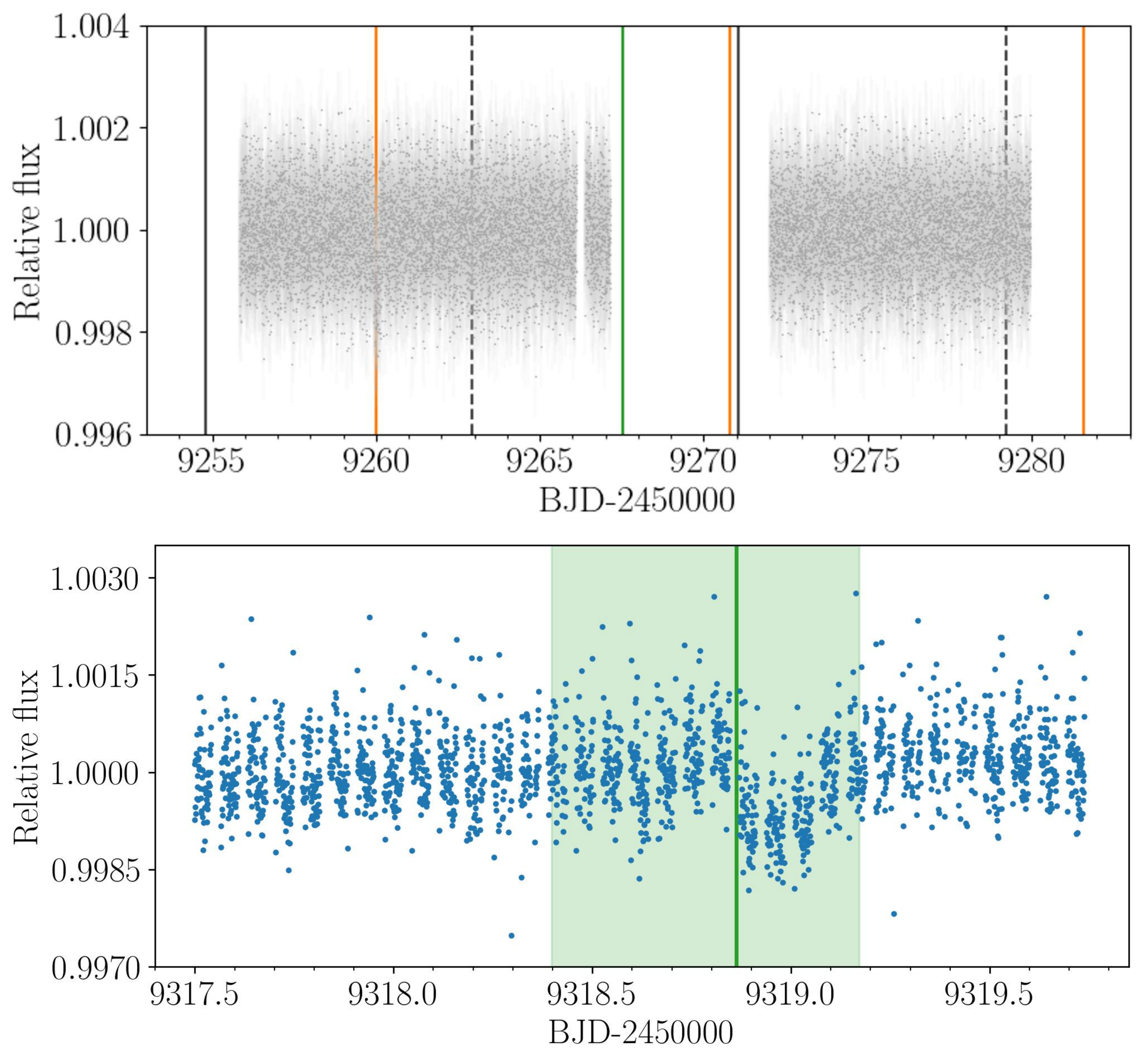}
  \caption{{\it Top}: 2-min cadence detrended \tess\ light curve of sector 35. The predicted transit times of TOI-561 c and d (according to L21 ephemeris), and TOI-561 f (according to W21 ephemeris), are highlighted with orange, green and black vertical solid lines, respectively. The black dashed lines indicate the predicted position of planet f alias at $\sim 8.1$~d. The only transit present in the light curve is the one of TOI-561 c at $\sim 9260$~ BJD$-2450000$, while no transit events occurred at the predicted times of planet f alias. The transits of planet d and f fall into the time series gap. 
  {\it Bottom}: \cheops\ visit scheduled to observe TOI-561 d. The green vertical solid line indicates the predicted transit time used to compute the \cheops\ observing window after the ephemeris update (Section~\ref{sec:pl_d}). The transit occurred within the 68 per cent highest probability density interval, highlighted by the pale green region. We note that this transit is not consistent with the ephemeris propagation of planet f, which would have transited at $9319.94$ BJD$-2450000$, so almost one day after the observed \cheops\ transit.
    }
    \label{fig:ephemeris}
\end{figure}

{\label{sec:pl_d}}
To solve the discrepancy among the planetary architectures proposed by L21 and W21 (Section~\ref{sec:pl_system}), we initially looked for the transits of TOI-561 d ($\sim 25$~d) and TOI-561 f ($\sim 16$~d) in the \tess\ sector $35$ light curve, whereas TOI-561 e ($\sim 77$~d) was not expected to transit during those \tess\ observations.
However, as shown in the top panel of Figure~\ref{fig:ephemeris}, the transits of planet d and f occurred during the light curve gap (Section~\ref{sec:tess_observations}), and so we could not use the new \tess\ data to conclusively discriminate between the two planetary configurations.
Nonetheless, these observations ruled out the planet f alias at $\sim 8.1$~d mentioned in W21, since no transit events were detected at its predicted transit times. 

We therefore decided to probe the L21 scenario collecting a transit of TOI-561 d using \cheops. 
We opted for the scheduling of the last seasonal observing window, in April, in order to take advantage of the most updated ephemeris to optimize the scheduling. 
For this reason, we performed a global fit adding to the literature data a partial set of the new HARPS-N RVs, as of 16 March 2021, and including the \tess\ sector $35$ light curve extracted from the second data release of the \texttt{tica} FFIs in March 2021. 
Even if no transit was detected, the new \tess\ sector helped to reduce the time window in which to search. 
In fact, the \tess\ data partially covered the $3 \sigma$-uncertainty transit window, enabling us to exclude some time-spans in the computation of the \cheops\ visit.
Thanks to the ephemeris update, the  \cheops\ $3 \sigma$ observing window shrank from $\sim 7.4$~d to $\sim 2.2$~d, demonstrating the importance of the early \tess\ data releases in the scheduling of follow-up observations. 
The bottom panel of Figure~\ref{fig:ephemeris} shows the \cheops\ visit scheduled to observe TOI-561 d, whose transit occurred almost exactly at the predicted time, so confirming the planetary period and giving further credence to the 4-planet scenario proposed by L21 .

Even updating the ephemeris using the partial new HARPS-N data set, the last possible \cheops\ observing window of TOI-561 e in the 2021 season was still longer than seven days because of ephemeris uncertainties. 
Even including the full set of RVs would have not helped as the target was no longer observable with \cheops\ when the HARPS-N campaign finished.
Given the high pressure on the \cheops\ schedule, we therefore plan the TOI-561 e observations for the 2022 observing season.
The ephemeris for the 2022 \cheops\ observations will be updated using the \tess\ Sectors 45 and 46 observations in Nov-Dec 2021, and the results will be presented in a future publication.

\subsection{Additional signals in the RV data}{\label{sec:additional_pl}}

\begin{figure}
\centering
  \includegraphics[width=\linewidth]{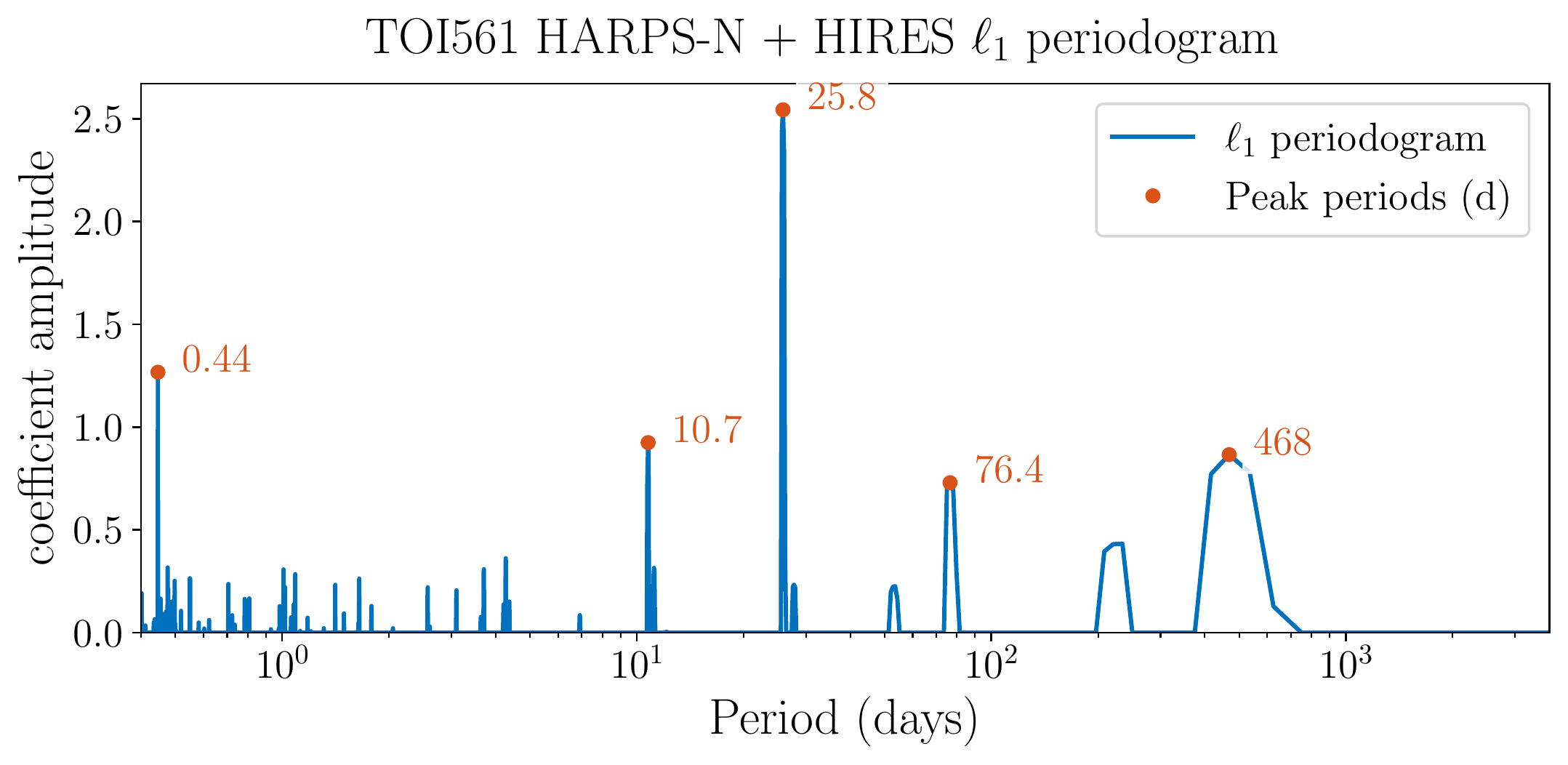}
  \caption{$\ell_1$-periodogram of the combined HARPS-N and HIRES data sets, computed on a grid of frequencies from $0$ to $2.5$ cycles per day. The total time-span of the observations is $768$~days. The code automatically accounts for the offset between HARPS-N and HIRES data by using the mean value of each data set.
    }
    \label{fig:l1_periodogram}
\end{figure}

\begin{figure}
\centering
  \includegraphics[width=\linewidth]{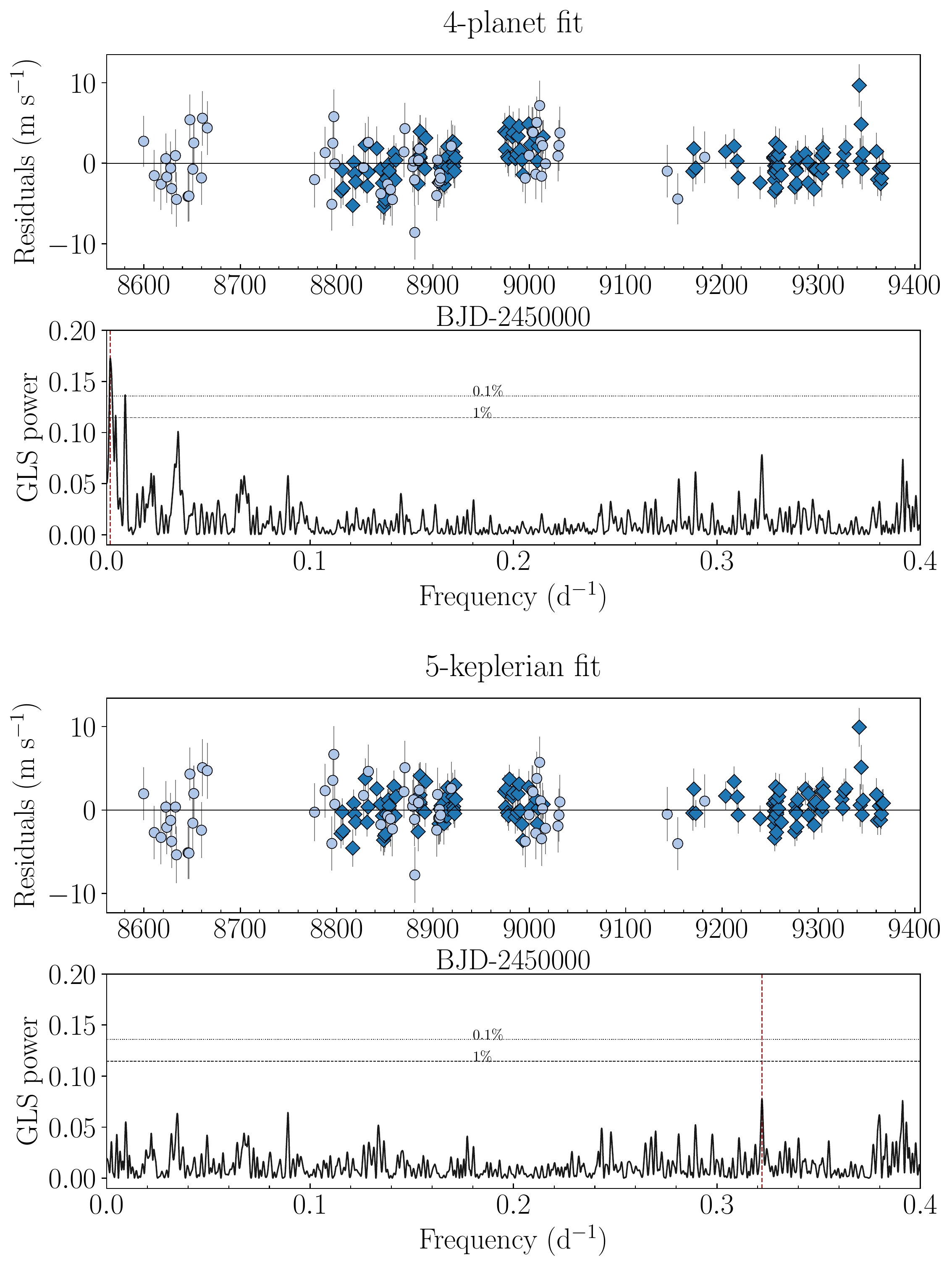}
  \caption{Time series and  GLS periodogram of the RV residuals after the 4-planet and 5-Keplerian fits as described in Section~\ref{sec:additional_pl}. 
  In the residuals plot, the HARPS-N and HIRES RVs are plotted with dark blue diamonds and light blue circles, respectively. 
  In the periodogram plots, the dashed and dotted horizontal lines show the $1$ and $0.1$ per cent False Alarm Probability (FAP) level, respectively. The red vertical line indicates the main peak of each periodogram. 
  The long-period peak around frequencies $0.0017-0.0025$~d$^{-1}$ ($P = 400-600$~d) in the 4-planet residuals periodogram is modelled by the fifth Keplerian, and no more significant peaks are identified in the 5-Keplerian residuals periodogram. Moreover, the positive variation at $\sim 9000$~BJD-2450000 in the 4-planet fit residuals disappears in the 5-Keplerian fit residuals.
    }
    \label{fig:res_4pl}
\end{figure}

Before proceeding with the global modelling, we analyzed the RV data sets in order to confirm the robustness of the L21 scenario and search for potential new signals. 
The $\ell_1$-periodogram\footnote{\url{https://github.com/nathanchara/l1periodogram}.} \citep{hara2017} of the combined HARPS-N and HIRES RVs (Figure~\ref{fig:l1_periodogram}) shows four significant peaks corresponding to the planetary periods reported in L21, plus hints of a possible longer period signal with a broad peak around $400-600$~days.
We investigated the presence of this additional signal in a Bayesian framework using  \texttt{PyORBIT}\footnote{\url{https://github.com/LucaMalavolta/PyORBIT}, V8.1.} \citep{Malavolta2016, Malavolta2018}, 
a package for light curve and RVs analysis.
We employed the \texttt{dynesty} nested-sampling algorithm \citep{skilling2004, skilling2006, speagle2020}, assuming $1000$ live points, and including offset and jitter terms for each data set. 
We first performed a 4-planet fit of the combined data sets, using the L21 values to impose Gaussian priors on periods and transit times,\footnote{We note that we obtained the same results when using uniform, uninformative priors, also for the 5- and 6-Keplerian fits.} and assuming eccentric orbits with a half-Gaussian zero-mean prior on the eccentricity (with variance $0.098$; \citealt{vanEylen2019}), except for the circular orbit of the USP planet. 
We let the semi-amplitude $K$  vary between $0.01$ and $100$ \ms.
As can be seen in Figure~\ref{fig:res_4pl}, the RV residuals show an anomalous positive variation at $\sim 9000$~BJD-$2450000$, and the Generalized Lomb-Scargle (GLS, \citealt{Zechmeister2009}) periodogram of the RV residuals revealed the presence of a significant, broad peak at low frequencies. 
Moreover, the HARPS-N jitter was $1.84$~\ms, which is unusually high when compared to the value reported in L21 ($\sigma_{\rm HARPS-N} = 1.29 \pm 0.23$~\ms). 
We therefore performed a second fit including a fifth Keplerian signal, allowing the period to span between $2$ and $900$~d.
According to the Bayesian Evidence, this model is strongly favoured with respect to the 4-planet model, with a difference in the logarithmic evidences $ \Delta\ln\mathcal{Z} = 19.0$ \citep{kass&raftey1995}.\footnote{According to \citet{kass&raftey1995}, a difference $\Delta\ln\mathcal{Z} > 5$ sets a strong evidence against the null hypothesis, which in our case corresponds to the 4-planet model.}
Moreover, the HARPS-N jitter decreased to $\sim 1.37$~\ms.
After this fit, the periodogram of the residuals did not show evidence of additional significant peaks (Figure~\ref{fig:res_4pl}).
This is confirmed also by the comparison with a 6-Keplerian model that we tested, with the period of the sixth Keplerian free to span between $2$ and $900$~d, whose Bayesian Evidence differed by less than $2$ from the 5-Keplerian model one, indicating that there was no strong evidence to favour a more complex model \citep{kass&raftey1995}. 

\begin{figure}
\centering
  \includegraphics[width=\linewidth]{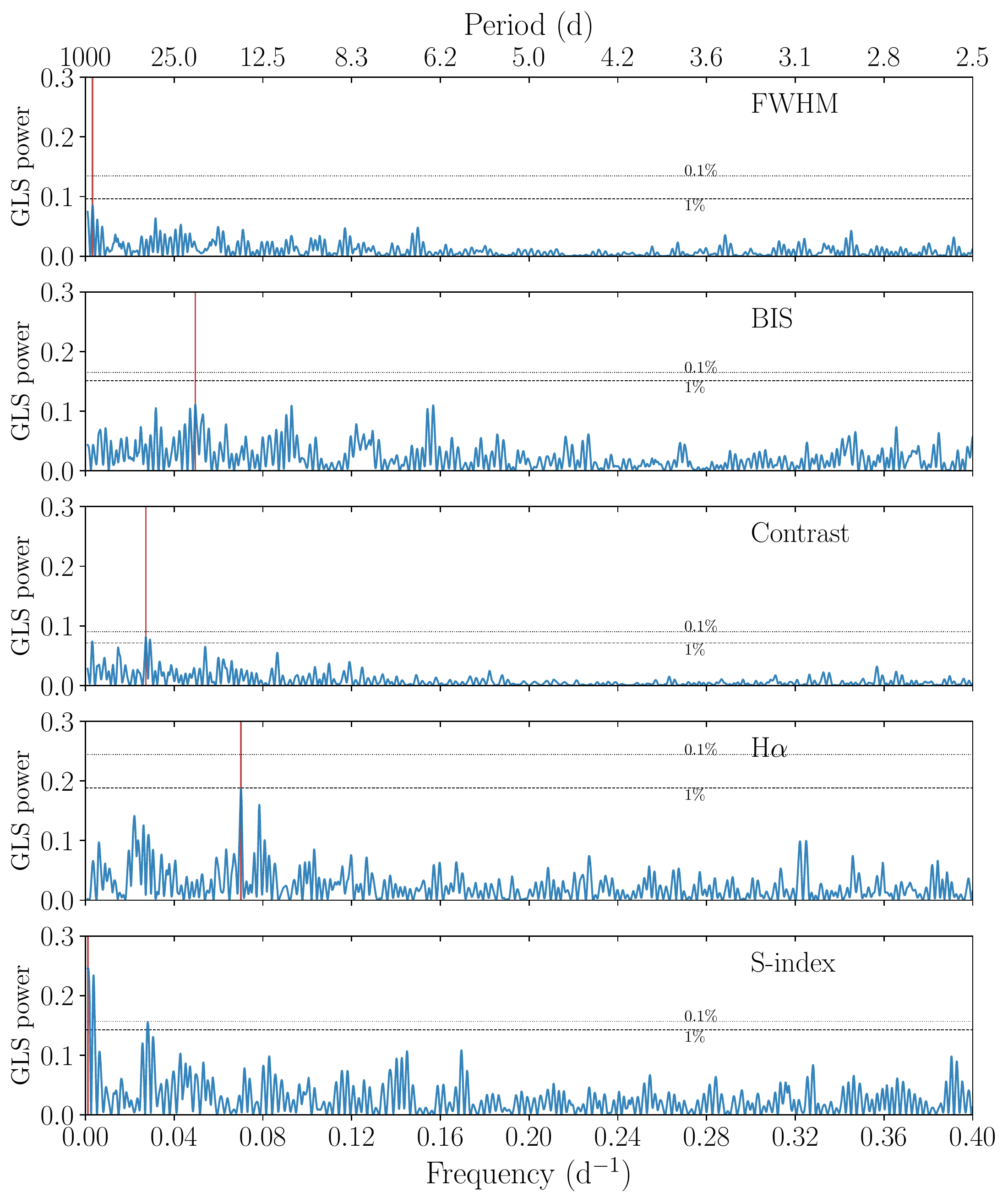}
  \caption{GLS periodogram of the HARPS-N activity indices. The main peak of each periodogram is highlighted with a red vertical line. The dashed and dotted horizontal lines indicate the $1$ and $0.1$ per cent FAP levels, respectively.
  The only peak above the $0.1$ FAP level is the low-frequency peak in the S-index periodogram, as discussed in Section~\ref{sec:additional_pl}.
    }
    \label{fig:periodograms_activity}
\end{figure}

The fitted period of the fifth Keplerian was $\sim 480$~d. 
Such a long-term signal could be induced either by stellar activity, considering that stellar magnetic fields related to magnetic cycles can show variability on timescales of the order of $1-3$ years (e.g. \citealt{collier_cameron_2018, hatzes_2019, eprv2021}), or by an additional long-period planet.
We refer here to an eventual long-period planet because, given the inferred semi-amplitude of $\sim 2$~\ms\ (Table~\ref{table:joint_parameters}), an external companion with mass equal to $13$~$M_{\rm j}$ (assuming this value to be the threshold between planetary and sub-stellar objects) would have an inclination of $\sim 0.03$~deg. Such an inclination would imply an almost perpendicular orbit with respect to the orbital plane of the four inner planets, hinting at a very unlikely configuration. Therefore, in the hypothesis of the presence of an external companion, it would most likely be a planetary-mass object. \\
On one hand, all the five signals, including the long-term one, are recovered in an independent analysis that we performed with the CCF-based \textsc{scalpels} algorithm \citep{collier_cameron_2021}.
Concisely, \textsc{scalpels} projects the RV time series onto the highest variance principal components of the time series of autocorrelation functions of the CCF, with the aim of distinguishing RV variations caused by orbiting planets from activity-induced distortions on each CCF.
The absence of the signal in the {\sc scalpels} shape-driven velocities indicates that the long-term periodicity is not due to shape changes in the line profiles, supporting the idea of a planetary origin.
Moreover, TOI-561 is not expected to be a particularly active star given its old age and low \logRHK, as assessed in the L21 and W21 activity analyses.
As can be seen in Figure~\ref{fig:periodograms_activity}, the GLS periodogram of the majority of the activity indicators extracted with the HARPS-N DRS, i.e. full width at half maximum (FWHM), bisector span (BIS), contrast and H$_{\alpha}$, do not show significant peaks, with none of them exceeding the $0.1$ False Alarm Probability (FAP) threshold, which we computed using a bootstrap approach, at the frequency of interest.
On the other hand, the periodogram of the S-index, which is particularly sensitive to magnetically-induced activity, shows a significant, broad peak at low frequencies, potentially suggesting that the previously identified long-term variability is related to stellar activity. Considering this, we performed an additional \texttt{dynesty} fit assuming a 4-planet model and including a Gaussian Process (GP) regression with a quasi-periodic kernel, as formulated in \citet{Grunblatt2015}, to account for the long-term signal. 
We modelled simultaneously the RVs and the S-index time series in order to better inform the GP \citep{langellier2021, osborn_hugh2021}, using two independent covariance matrices for each dataset with common GP hyper-parameters except for the amplitude of the covariance matrix, assuming uniform, non-informative priors on all of them.
The fit suggests a periodicity longer than $\sim 570$~d, but the GP model is too flexible to derive a precise period value, considering also that the global RV baseline ($\sim 768$~d) is comparable with the periodicity of the long-term signal. 
The inferred semi-amplitudes of the four known planets differed by less than $0.07 \sigma$ from the 5-Keplerian model ones, indicating that the different modelling of the long-term signal is not influencing the results for the known, transiting planets.
Finally, as in the case of the 5-Keplerian fit, the HARPS-N jitter is significantly improved ($\sigma_{\rm HARPS-N} \sim 1.30$ \ms) when including the GP model. 
Therefore, since our Bayesian analyses showed that the modelling of the long-term signal is necessary to obtain the best picture of the system, we decided to perform the global fit assuming a 5-Keplerian model, but without drawing conclusions on the origin of the fifth signal. 
We stress that the 5-Keplerian fit does not provide absolute evidence of the presence of a fifth planet, since also poorly sampled stellar activity could be well modelled using a Keplerian \citep{Pepe2013, mortier2017, affer2016}, especially in our case where the RV baseline is of the order of the signal periodicity.
Since it is not possible to distinguish a true planetary signal from an activity signal that has not been observed long enough to exhibit a loss of coherence in its phase or amplitude, only a follow-up campaign over several years can allow one to better understand the nature of this long-term signal.

\section{Joint photometric and RV analysis}{\label{sec:global_analysis}}

To infer the properties of the TOI-561 planets, we jointly modelled all photometric and spectroscopic data with \texttt{PyORBIT}, using \texttt{PyDE}\footnote{\url{https://github.com/hpparvi/PyDE}.} + \texttt{emcee} \citep{ForemanMackey2013} as described in Section 5 of L21, and adopting the same convergence criteria. We ran $96$ chains (twice the number of the model parameters) for $250000$ steps, discarding the first $50000$ as burn-in.

Based on the analysis presented in the previous section, we assumed a 5-Keplerian model, including four planets  plus a fifth Keplerian with period free to span between $2$ to $900$~d.
We fitted a common value for the stellar density, using the value reported in Table~\ref{table:star_params} as Gaussian prior. 
We adopted the quadratic limb-darkening law as parametrized by \citet{Kipping2013}, putting Gaussian priors on the $u_1$, $u_2$ coefficients, obtained for the \tess\ and \cheops\ passband through a bilinear interpolation of limb darkening profiles by \citet{claret2017} and \citet{claret2021} respectively, and assuming a $1 \sigma$ uncertainty of $0.1$ for each coefficient.
We imposed a half-Gaussian zero-mean prior \citep{vanEylen2019} on the planet eccentricities, except for the USP planet, whose eccentricity was fixed to zero. 
We assumed uniform priors for all the other parameters.

To model the long-term correlated noise in the \tess\ light curve, we included in the fit a GP regression with a Matérn-3/2 kernel against time, as shown in Figure~\ref{fig:tess_lc}, and we added a jitter term to account for possible extra white noise.
We pre-decorrelated the \cheops\ light curves with the \texttt{pycheops}\footnote{\url{https://github.com/pmaxted/pycheops}.} package \citep{maxted2021}, selecting the detrending parameters according to the Bayes factor to obtain the best correlated noise model for each visit. 
For all the three \cheops\ visits, a decorrelation for the first three harmonics of the roll angle was necessary, plus first-order polynomials in time, x-y centroid position, and smearing. 
We then used the detrended light curves (Figure~\ref{fig:cheops_lc}) for the global \texttt{PyORBIT} fit.
In order to check if the detrending was affecting our results for the planetary parameters, we also performed an independent global analysis with the \texttt{juliet} package \citep{espinoza2019}, including in the global modelling the basis vectors selected with \texttt{pycheops} to detrend the data simultaneously. 
All the results were consistent within $1 \sigma$,
indicating that the pre-detrending did not significantly alter our inferred results.
Finally, for both the HARPS-N and HIRES data sets we included jitter and offset terms as free parameters. 

\begin{figure}
\centering
  \includegraphics[width=\linewidth]{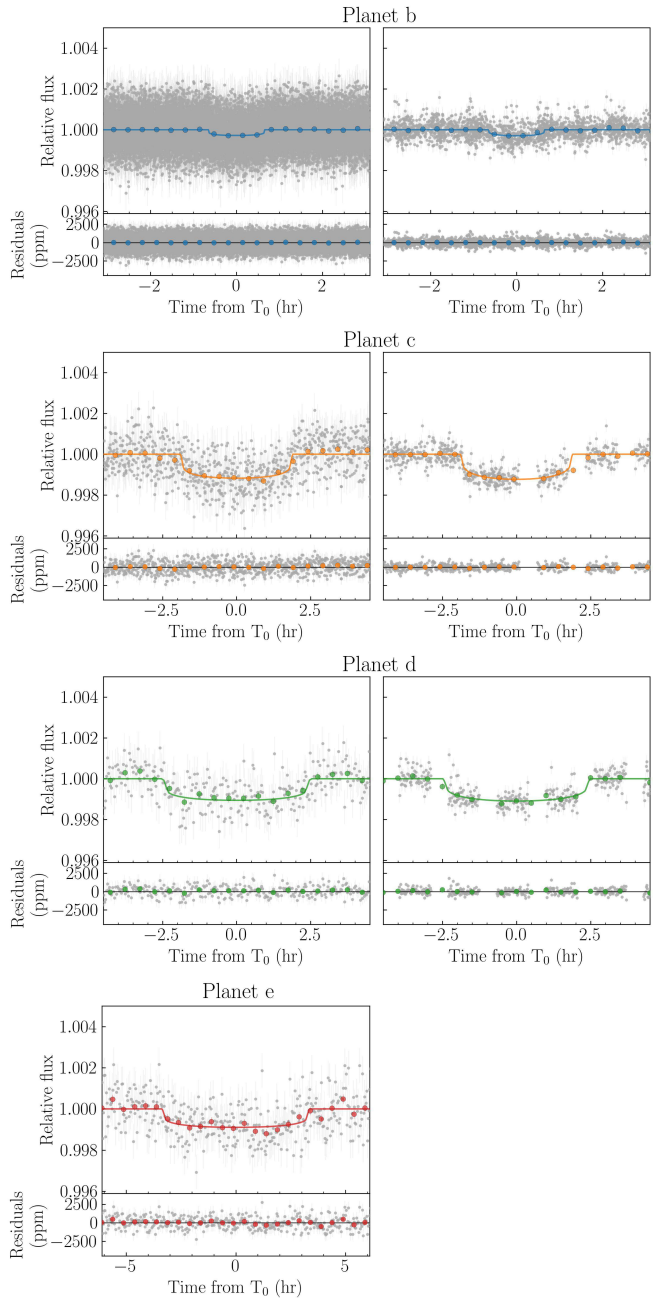}
  \caption{Phase-folded \tess\ ({\it left}) and \cheops\ ({\it right}) light curves of TOI-561 b, c, and d. Planet e shows a single transit in the \tess\ light curve,  and it has no \cheops\ observations. For each planet, the coloured line indicates the best-fitting model, and residuals are shown in the bottom panels. Data points binned over $20$~min (planet b) and $30$~min (planets c, d and e) are shown with coloured dots. 
    }
    \label{fig:transit_model}
\end{figure}

\begin{figure}
\centering
  \includegraphics[width=\linewidth]{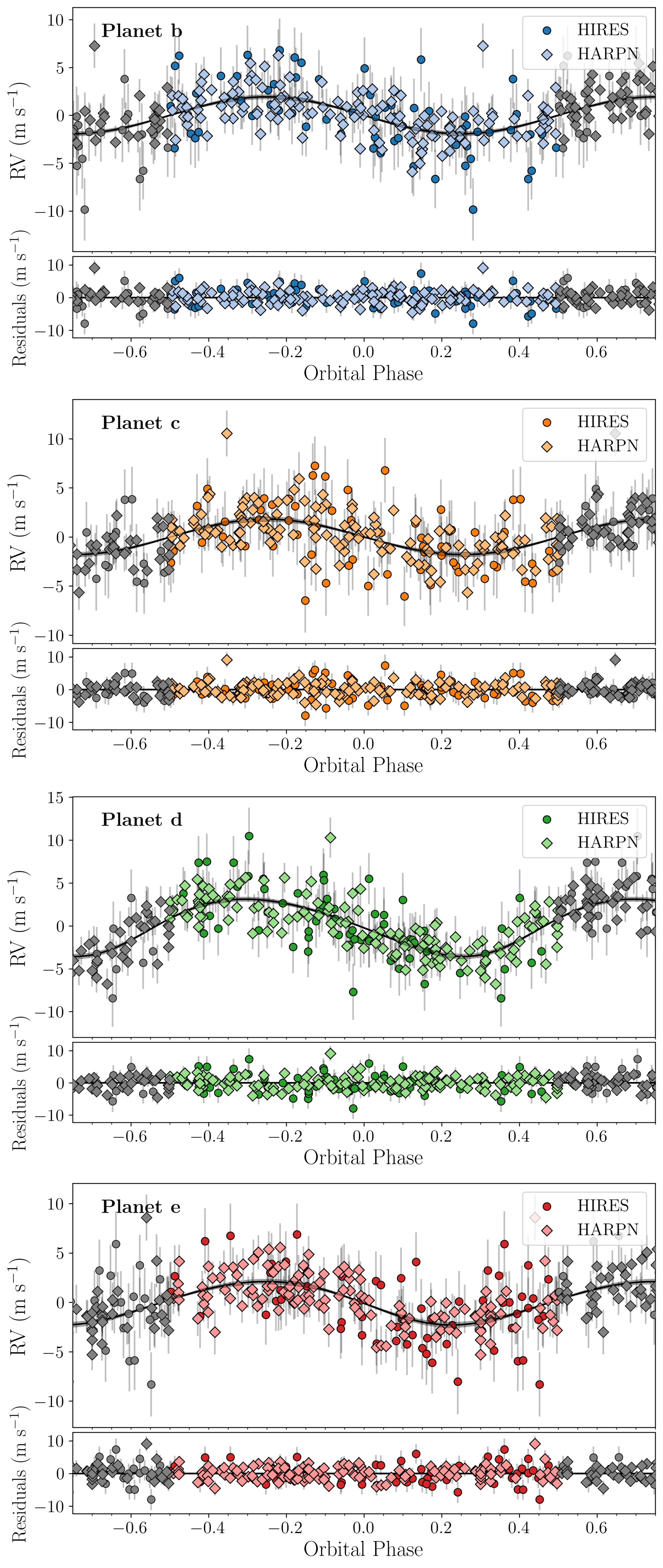}
  \caption{Phase-folded HARPS-N and HIRES RVs with residuals of TOI-561 b, c, d and e, as resulting from the joint photometric and spectroscopic fit. The error bars include the jitter term added in quadrature. 
    }
    \label{fig:folded_RV}
\end{figure}

\begin{figure}
\centering
  \includegraphics[width=\linewidth]{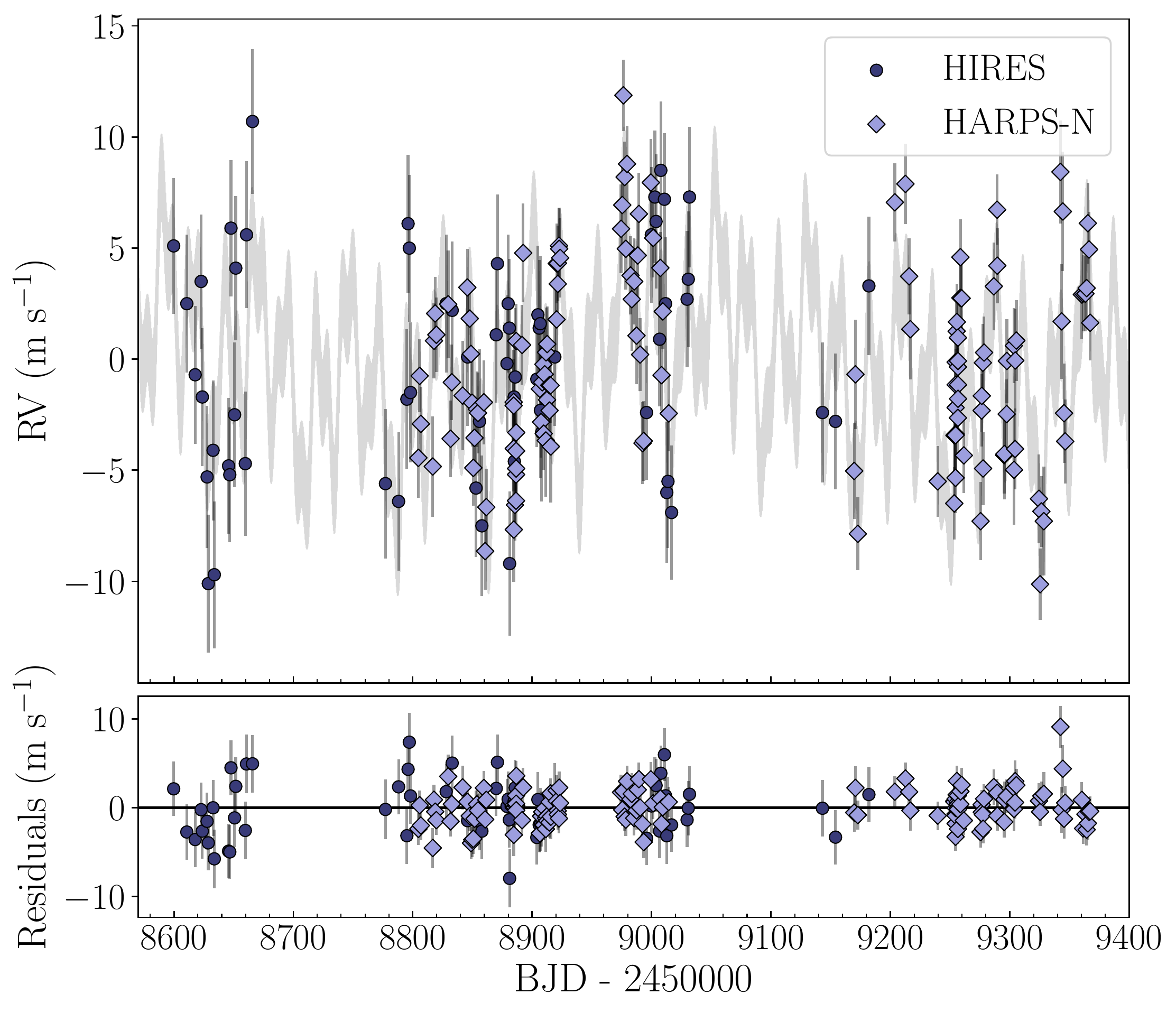}
  \caption{Global model (grey line) with residuals of HARPS-N and HIRES RVs according to the 5-Keplerian photometric and spectroscopic fit. The error bars include the jitter term added in quadrature. 
    }
    \label{fig:global_RV}
\end{figure}

\begin{figure}
\centering
  \includegraphics[width=\linewidth]{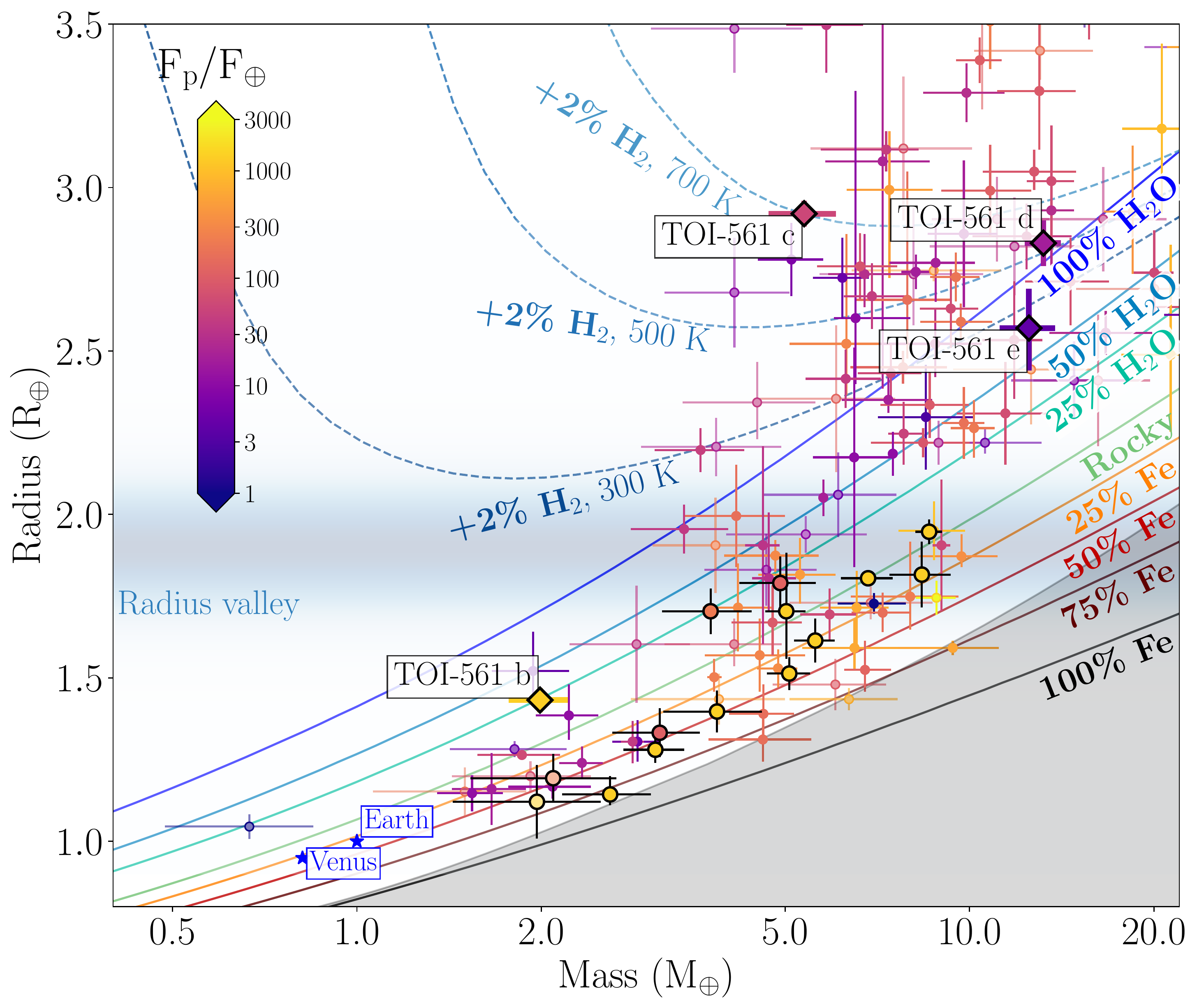}
  \caption{Mass-radius diagram for exoplanets with radii and masses measured with a precision better than $30$\%, colour coded according to their incidental flux. Data are taken from the Extrasolar Planets Encyclopaedia catalogue (\url{http://exoplanet.eu/catalog/}) as of 18 October 2021. The TOI-561 planets are labelled, and highlighted with coloured diamonds. The USP planets are emphasized with thick, black-contoured circles. The theoretical mass-radius curves for various chemical compositions \citep{Zeng2019} are represented by solid coloured lines, while the dashed lines indicate the curves for an Earth-like core surrounded by a H$_2$ envelope ($2$\% mass fraction) at varying equilibrium temperatures. The forbidden region predicted by collisional stripping \citep{marcus2010} is marked by the shaded grey region.
    }
    \label{fig:MR_diagram}
\end{figure}

\begin{figure}
\centering
  \includegraphics[width=\linewidth]{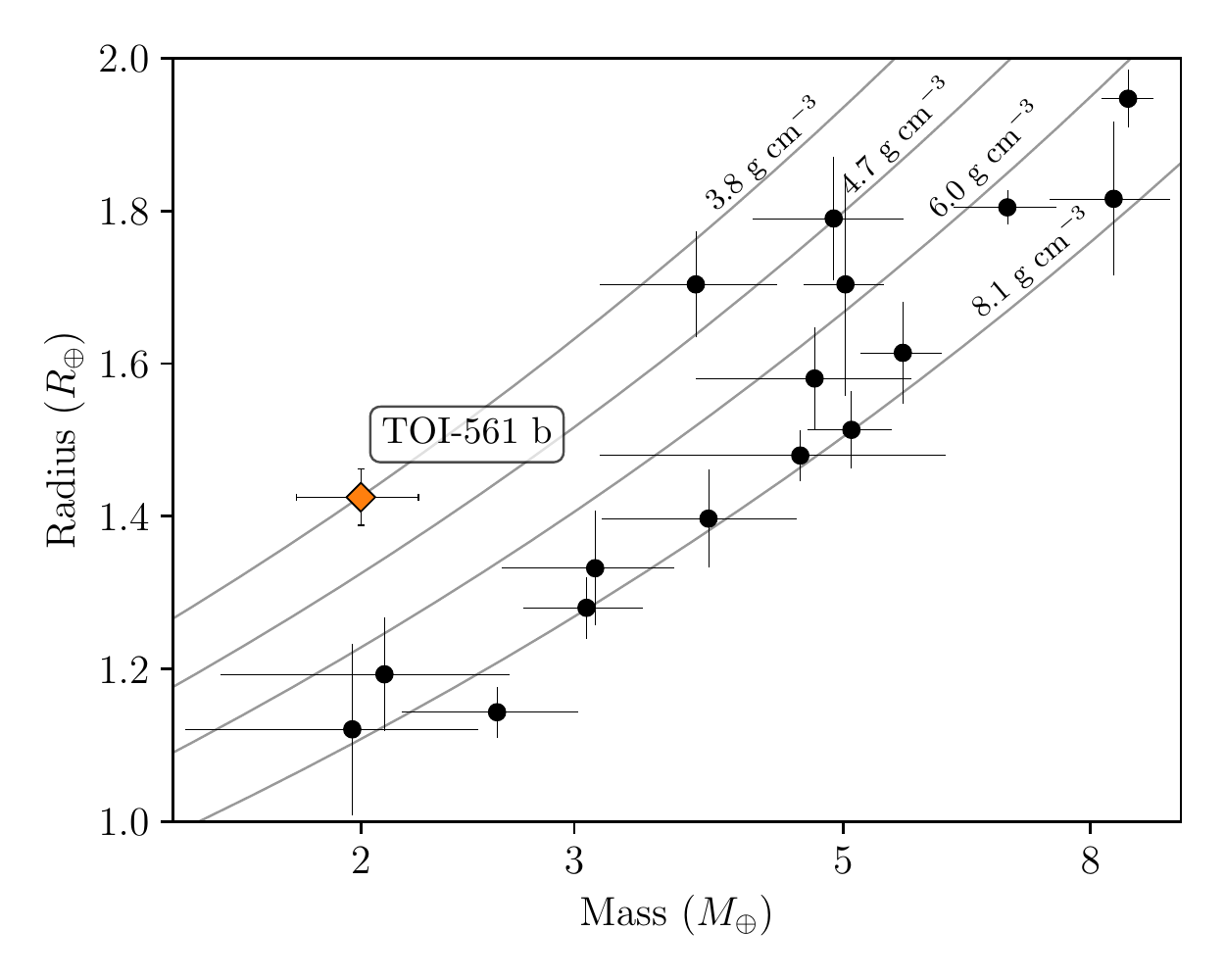}
  \caption{Mass-radius diagram of confirmed USP planets ($P < 1$~d, \rplanet < 2~\rearth) as taken from the Extrasolar Planets Encyclopaedia catalogue in date 18 October 2021. 
  Iso-density lines are plotted in grey. TOI-561 b stands out as the lowest density USP planet known to date ($\rho_{\rm b} = 3.8 \pm 0.5$ \gcm).
    }
    \label{fig:USPs_MR}
\end{figure}

We summarize our best-fitting model results in Table~\ref{table:joint_parameters}, and we show the transit model, phase-folded RVs and global RV model in Figures~\ref{fig:transit_model}, \ref{fig:folded_RV}, and \ref{fig:global_RV}, respectively. 
We inferred precise masses and radii for all the four planets in the system, whose positions in the mass-radius diagram are shown in Figure~\ref{fig:MR_diagram}.
With a radius of $R_{\rm b} = 1.425 \pm 0.037$~\rearth\ and a mass of $M_{\rm b} = 2.00 \pm 0.23$~\mearth\ (from $K_{\rm b} = 1.93 \pm 0.21$~\ms), TOI-561 b is located in a region of the mass-radius diagram which is not consistent with a pure rocky composition, as will be also shown in Section~\ref{sec:internal_model} by our internal structure modelling. Our analysis confirms TOI-561 b to be the lowest density ($\rho_{\rm b} = 3.8 \pm 0.5$ \gcm) USP planet known to date (Figure~\ref{fig:USPs_MR}).
In order to further confirm the planetary density, we also performed a specific RV analysis of TOI-561 b using the Floating Chunk Offset method (FCO; \citealt{Hatzes2014}). 
The FCO analysis, detailed in Appendix~\ref{sec:app_FCO}, confirms the low mass inferred for TOI-561 b, and consequently its low density.
Thanks to the \cheops\ observations, we also improved significantly the radius of TOI-561 c, for which we obtained a value of $R_{\rm c} = 2.91 \pm 0.04$~\rearth. 
From the semi-amplitude $K_{\rm c} = 1.81_{-0.22}^{+0.23}$~\ms\ we inferred a mass of $M_{\rm c} = 5.39_{-0.68}^{+0.69}$~\mearth, implying a density of $\rho_{\rm c} = 1.2 \pm 0.2$~\gcm.
From the combined fit of one \tess\ and one \cheops\ transit, we inferred a radius of $2.82 \pm 0.07$~\rearth\ for planet d, which has a mass of $M_{\rm c} = 13.2_{-0.9}^{+1.0}$~\mearth\ (from $K_{\rm d} = 3.34_{-0.22}^{+0.23}$~\ms) and a resulting density of $\rho_{\rm d} = 3.2 \pm 0.3$~\gcm. 
Finally, for TOI-561 e, which shows a single transit in {\it TESS} sector $8$, we derived a radius of $R_{\rm e} = 2.55_{-0.13}^{+0.12}$~\rearth, a mass of $M_{\rm e} = 12.6 \pm 1.4$~\mearth, and an average density of $\rho_{\rm d} = 4.2 \pm 0.8$~\gcm. 
Lastly, the period inferred for the fifth Keplerian in the model was $473_{-25}^{+36}$~d, with a $7.2\sigma$ detected semi-amplitude of $1.94 \pm 0.27$~\ms. 
As discussed in Section~\ref{sec:additional_pl}, additional data spanning a longer baseline are needed to definitively confirm the planetary nature of this long-term signal.

\begin{table*}
\caption{Parameters of the TOI-561 system, including the fifth Keplerian signal, as determined from the joint photometric and spectroscopic fit described in Section~\ref{sec:global_analysis}.}
\label{table:joint_parameters} 
\begin{threeparttable}[t]
\centering
\begin{tabular}{l c c c c c} 
  \hline\hline       
 \multicolumn{6}{c}{Planetary parameters}\\[1ex]
 \hline 
   & TOI-561 b & TOI-561 c & TOI-561 d & TOI-561 e & $5^{\rm th}$ Keplerian\\
  \hline
  $P$ (d) & $0.4465688_{-0.0000008}^{+0.0000007}$& $10.778831_{-0.000036}^{+0.000034}$& $25.7124_{-0.0002}^{+0.0001}$ & $77.03_{-0.24}^{+0.25}$ & $473_{-25}^{+36}$ \\
  $T_0$ (TBJD)$^a$ & $2317.7498 \pm 0.0005$ & $2238.4629_{-0.0009}^{+0.0008}$  & $2318.966_{-0.004}^{+0.003}$ & $1538.180_{-0.005}^{+0.004}$ & $1664_{-33}^{+28}$\\
  $a/$\rstar & $2.685_{-0.025}^{+0.024}$ & $22.43_{-0.21}^{+0.20}$& $40.04_{-0.37}^{+0.36}$ & $83.22_{-0.79}^{+0.77}$ & $279_{-10}^{+14} $ \\
  $a$ (AU)  & $0.0106 \pm 0.0001$ & $0.0884 \pm 0.0009$ & $0.158 \pm 0.002$ & $ 0.328 \pm 0.003$ & $1.1_{-0.4}^{+0.6}$\\
  $R_\mathrm{p}/$\rstar & $0.0155 \pm 0.0004$ & $0.0316 \pm 0.0004$& $0.0306 \pm 0.0008$ &  $0.0278_{-0.0014}^{+0.0016}$ & - \\
  \rplanet\ (\rearth) & $1.425 \pm 0.037$& $2.91 \pm 0.04$ & $2.82 \pm 0.07$ & $2.55_{-0.13}^{+0.12}$ & - \\
  $b$ & $0.13_{-0.09}^{+0.10}$ & $0.12_{-0.08}^{+0.13}$ & $0.45_{-0.17}^{+0.11}$ & $0.28_{-0.18}^{+0.15}$ & -\\
  $i$ (deg) & $87.2_{-2.1}^{+1.9}$ & $89.69_{-0.31}^{+0.21}$ & $89.40_{-0.11}^{+0.21}$ & $89.80_{-0.10}^{+0.13}$ & -\\
  $T_{14}$ (h) & $1.31 \pm 0.02$ & $3.75_{-0.08}^{+0.05}$& $4.54_{-0.29}^{+0.32}$ & $6.98_{-0.40}^{+0.24}$ & -\\
  $e$ & $0$ (fixed) & $0.030_{-0.021}^{+0.035}$& $0.122_{-0.048}^{+0.054}$ & $0.079_{-0.050}^{+0.058}$ & $0.085_{-0.059}^{+0.083}$ \\
  $\omega$ (deg) & $90$ (fixed) & $291_{-84}^{+55}$ & $235_{-26}^{+14}$ & $143_{-44}^{+42}$ & $348_{-53}^{+198}$\\
  $K$ (\ms) &$1.93 \pm 0.21$ & $1.81_{-0.22}^{+0.23}$& $3.34_{-0.22}^{+0.23}$ & $2.19 \pm 0.23$ & $1.94 \pm 0.27$\\
  \mplanet\ (\mearth) &$2.00 \pm 0.23$ & $5.39_{-0.68}^{+0.69}$& $13.2_{-0.9}^{+1.0}$ & $12.6 \pm 1.4$ & $20 \pm 3 \, ^b$\\
  \rhoplanet\ (\rhoearth) & $0.69 \pm 0.10$& $0.22 \pm 0.03$& $ 0.59 \pm 0.06$& $0.76 \pm 0.14$ & -\\
  \rhoplanet\ (\gcm) & $3.8 \pm 0.5$ & $1.2 \pm 0.2$& $3.2 \pm 0.3$ & $4.2 \pm 0.8$ & -\\
  $S_{\rm p}$ ($S_{\oplus}$) & $4745 \pm 269$ & $68.2 \pm 3.9$& $21.4 \pm 1.3$ & $4.96 \pm 0.28$ & -\\
  $T_{\rm eq}^c$ (K) & $2310 \pm 33$ & $800 \pm 11$ & $598 \pm 9$ & $415 \pm 6$ & -\\
  $g_{\rm p}^d$ (m s$^{-2}$) & $9.7 \pm 1.2$ & $6.2 \pm 0.8$ & $16.3 \pm 1.5$ & $19.0 \pm 2.9$ & -\\[1ex]
 \hline 
 \multicolumn{6}{c}{Common parameters}\\
 \hline
  \rstar$^{\rm e}$ (\rsun) & \multicolumn{4}{c}{$0.843 \pm 0.005$}\\
  \mstar$^{\rm e}$ (\msun) & \multicolumn{4}{c}{$0.806 \pm 0.036$}\\
  \rhostar\ (\rhosun) & \multicolumn{4}{c}{$1.31 \pm 0.05$} \\
  $u_{1, \mathrm{TESS}}$ & \multicolumn{4}{c}{$0.33 \pm 0.08$} \\
  $u_{2, \mathrm{TESS}}$ & \multicolumn{4}{c}{$0.23 \pm 0.09$} \\
  $u_{1, \mathrm{CHEOPS}}$ & \multicolumn{4}{c}{$0.46 \pm 0.07$} \\
  $u_{2, \mathrm{CHEOPS}}$ & \multicolumn{4}{c}{$0.22 \pm 0.09$} \\
  $\sigma_{\rm HARPS-N}^{\rm f}$ (\ms) & \multicolumn{4}{c}{$1.40_{-0.14}^{+0.15}$} \\
  $\sigma_{\rm HIRES}^{\rm f}$ (\ms) & \multicolumn{4}{c}{$2.77_{-0.31}^{+0.36}$} \\
  $\gamma^{\rm g}_{\rm HARPS-N}$ (\ms) & \multicolumn{4}{c}{$79700.41 \pm 0.26$} \\
  $\gamma^{\rm g}_{\rm HIRES}$ (\ms) & \multicolumn{4}{c}{$-1.20 \pm 0.42$} \\[1ex]
 \hline
\end{tabular}
\begin{tablenotes}
$^a$ TESS Barycentric Julian Date (BJD$-2457000$). $^b$ Minimum mass in the hypothesis of a planetary origin. $^c$ Computed as $T_{\rm eq} = T_\star \, \biggl(\dfrac{R_\star}{2a}\biggr)^{1/2} \, [f(1-A_{\rm B})]^{1/4}$, assuming $f=1$ and a null Bond albedo ($A_{\rm B} = 0$).  $^d$ Planetary surface gravity. $^e$ As determined from the stellar analysis in Section~\ref{sec:star}. $^f$ RV jitter term. $^g$ RV offset.
\end{tablenotes}
\end{threeparttable}
\end{table*}

\section{Internal structure modelling}{\label{sec:internal_model}}
We modelled the internal planetary structure in a Bayesian framework, following the procedure detailed in \cite{leleu2021}. 
Our model assumes fully-differentiated planets composed of four layers, comprising an iron and sulfur central core, a silicate mantle which includes Si, Mg and Fe, a water layer, and a pure H/He gas layer.
The inner core is modelled assuming the \cite{hakim2018} equation of state (EOS), the silicate mantle uses the \citet{sotin2007} EOS, and the water layer uses the \citet{haldemann2020} EOS.
The core, mantle and water layer compose the ‘solid’ part of the planet.
The thickness of the gas envelope is computed as a function of stellar age and irradiation, and mass and radius of the solid part, according to the model presented in \citet{Lopez_2014}. 
We assumed no compression effects of the gas envelope on the solid part, a hypothesis which is justified {\it a posteriori} given the low mass fraction of gas obtained for each planet (see below).

Our Bayesian model fits the planetary system as a whole, rather than performing an independent fit for each planet, in order to account for the correlations between the absolute planetary masses and radii, which depend on the stellar properties. 
The model fits the stellar (mass, radius, effective
temperature, age, chemical abundances of Fe, Mg, Si), and planetary properties (RV semi-amplitudes, transit depths, orbital periods) to derive the posterior distributions of the internal structure parameters. 
The internal structure parameters modelled for each planet are the mass fractions of the core, mantle and water layer, the mass of the gas envelope, the iron molar fraction in the core, the silicon and magnesium molar fraction in the mantle, the equilibrium temperature and the age of the planet (equal to the age of the star). For a more extensive discussion on the relation among input data and derived parameters we refer to \citet{leleu2021}.
We assumed the mass fraction of the inner core, mantle, and water layer to be uniform on the simplex (the surface on which they add up to one), with the water mass fraction having an upper boundary of 0.5 \citep{Thiabaud2014, marboeeuf2014}. 
For the mass of the gas envelope, we assumed a uniform prior in logarithmic space.
Finally, we assumed the Si/Mg/Fe molar ratios of each planet to be equal to the stellar atmospheric values (even though \cite{adibekyan2021} recently showed that the stellar and planetary abundances may not be always correlated in a one-to-one relation). 
We emphasize the fact that, as in many Bayesian analyses, the results presented below in terms of planet internal structure depend to some extent on the selection of the priors, which we chose following i.e. \citet{dorn2017}, \cite{dorn2018}, and \cite{leleu2021}.
Analysing the same data with very different priors (e.g. non uniform core/mantle/water mass fraction or gas fraction uniform in linear scale) would lead to different conclusions.

\begin{figure*}
\centering
  \includegraphics[width=\linewidth]{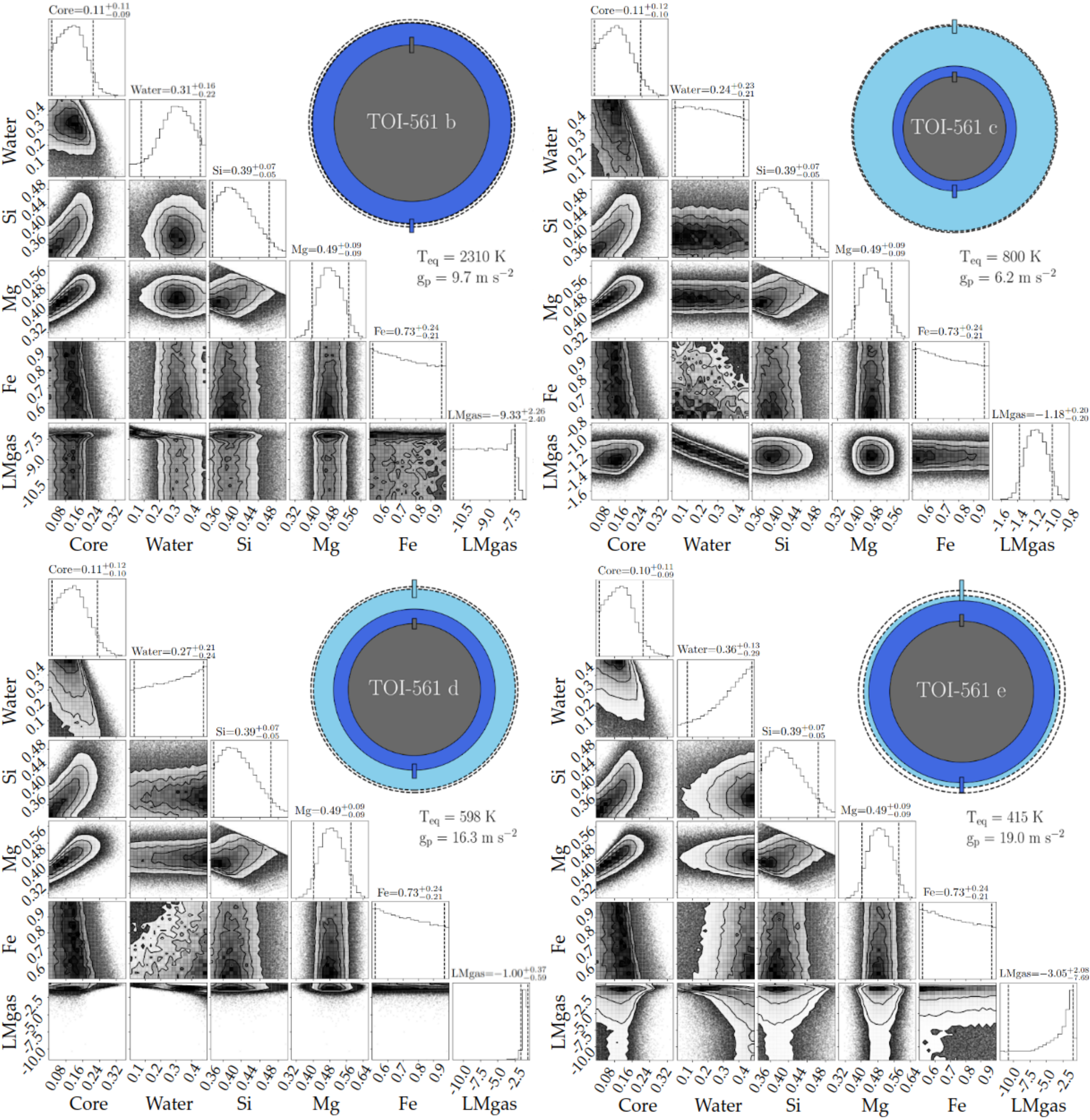}
  \caption{Posterior distributions of the main parameters describing the internal structure of TOI-561 b ({\it top left}), c ({\it top right}), d ({\it bottom left}), and e ({\it bottom right}). Each corner plot shows the mass fraction of the inner core and of the water layer, the molar fractions of silicon and magnesium in the mantle, the iron molar fraction in the inner core, and the mass of gas in logarithmic scale. On top of each column are printed the mean and the 5 per cent and 95 per cent quantiles values. For each planet, and we show an illustration of the radius fractions of the inner core+mantle (dark gray), water layer (dark blue), and gas envelope (light blue), corresponding to the medians of the posterior distributions. The coloured rectangles indicate the uncertainty on the corresponding layer thickness, while the black dashed outer rings represent the uncertainty on the total radius. Equilibrium temperature and planetary surface gravity are reported for each planet.
    }
    \label{fig:internal_model}
\end{figure*}

We show the results of the internal structure modelling for the four planets in Figure~\ref{fig:internal_model}.
As expected from its closeness to the host star, planet b has basically no H/He envelope, while the other three planets show a variable amount of gas mass.
Planet c hosts a relatively massive gaseous envelope, with a gas mass of (5 and 95 per cent quantiles) $M_{\rm gas,c} = 0.07_{-0.02}^{+0.04}$~\mearth\ ($1.3_{-0.4}^{+0.8}$ weight percent wt\%). 
Planet d hosts the most massive envelope ($M_{\rm gas, d} = 0.10_{-0.07}^{+0.13}$~\mearth), which, considering the total mass of the planet, correspond to a smaller relative mass fraction of $0.8_{-0.5}^{+1.0}$ wt\%, while TOI-561 e's envelope spans a range between $-10.7 < \log M_{\rm gas, e} < -1.0$, implying an upper limit on the gas mass of $0.11$~\mearth\ ($< 0.9$ wt\%). 
As expected from its low density, TOI-561 b could host a significant amount of water, having a water mass of $M_{\rm H_2O,b} = 0.62_{-0.44}^{+0.32}$~\mearth\ ($31_{-22}^{+16}$ wt\%). 
We stress that this result is highly dependent on the caveat of including only a solid water layer in the model. 
In fact, a massive water layer, if present on a planet with such a high equilibrium temperature, would imply the presence of a massive steam atmosphere \citep{turbet2020}. This would in turn considerably change the inferred water mass fraction with respect to a model that includes only a solid water layer.
Due to the presence of the gas envelope, the amount of water in both planet c and d is almost unconstrained ($M_{\rm H_2O,c} = 1.29_{-1.14}^{+1.24}$~\mearth, i.e. $24_{-21}^{+23}$ wt\%; $M_{\rm H_2O,d} = 3.56_{-3.18}^{+2.78}$~\mearth, i.e. $27_{-24}^{+21}$ wt\%), while TOI-561 e modelling points toward a massive water layer, with $M_{\rm H_2O,e} = 4.50_{-3.65}^{+1.69}$~\mearth\ ($36_{-29}^{+13}$ wt\%).

\section{Atmospheric evolution}{\label{sec:atmo_model}}

We employed the system parameters derived in this work to constrain the evolution of the stellar rotation period, which we use as a proxy for the evolution of the stellar high-energy emission affecting atmospheric escape, and the predicted initial atmospheric mass fraction of the detected transiting planets $f_{\rm atm}^{\rm start}$, that is the mass of the planetary atmosphere at the time of the dispersal of the protoplanetary disk. To this end, we used the planetary atmospheric evolution code \textsc{Pasta} described by \citet{bonfanti2021_atmo}, which is an updated version of the original code presented by \citet{kubyshkina2019a,kubyshkina2019b}. The code models the evolution of the planetary atmospheres combining a model predicting planetary atmospheric escape rates based on hydrodynamic simulations \citep[this has the advantage over other commonly used analytical estimates to account for both XUV-driven and core-powered mass loss;][]{kubyshkina2018}, a model of the stellar high-energy (X-ray plus extreme ultraviolet; XUV) flux evolution \citep{bonfanti2021_atmo}, a model relating planetary parameters and atmospheric mass \citep{johnstone15models}, and stellar evolutionary tracks \citep{Choi2016}. The main assumptions of the framework are that planet migration did not occur after the dispersal of the protoplanetary disk, and that the planets hosted at some point in the past or still host a hydrogen-dominated atmosphere.

For each planet, the evolution calculations begin at an age of 5\,Myr, which is the age assumed in the code for the dispersal of the protoplanetary disk. At each time step, the framework derives the mass-loss rate from the atmospheric escape model employing the stellar flux and the system parameters, and uses it to update the atmospheric mass fraction. This procedure is then repeated until the age of the system is reached or the planetary atmosphere has completely escaped. The free parameters of the algorithm are the initial atmospheric mass fraction at the time of the dispersal of the protoplanetary disk, and the indexes of the power law controlling the stellar rotation period \citep[see][for a detailed description of the mathematical formulation of the power law]{bonfanti2021_atmo}, that we use as proxy for the stellar XUV emission. 

The free parameters are constrained by implementing the atmospheric evolution algorithm in a Bayesian framework employing the MCMC tool presented by \citet{cubillos2017}. The framework uses the system parameters with their uncertainties as input priors. It then computes millions of forward planetary evolutionary tracks, varying the input parameters according to the shape of the prior distributions, and varying the free parameters within pre-defined ranges, fitting the current planetary atmospheric mass fractions obtained as described in Section~\ref{sec:internal_model}. The fit is done at the same time for all planets, thus simultaneously constraining the rotational period, and the results are posterior distributions of the free parameters. In particular, we opted for fitting for the planetary atmospheric mass fractions instead of the planetary radii. This enables the code to be more accurate by avoiding the continuous conversion of the atmospheric mass fraction into planetary radius, given the other system parameters \citep[see also][]{delrez2021}.

\begin{figure*}
\centering
  \includegraphics[width=\linewidth]{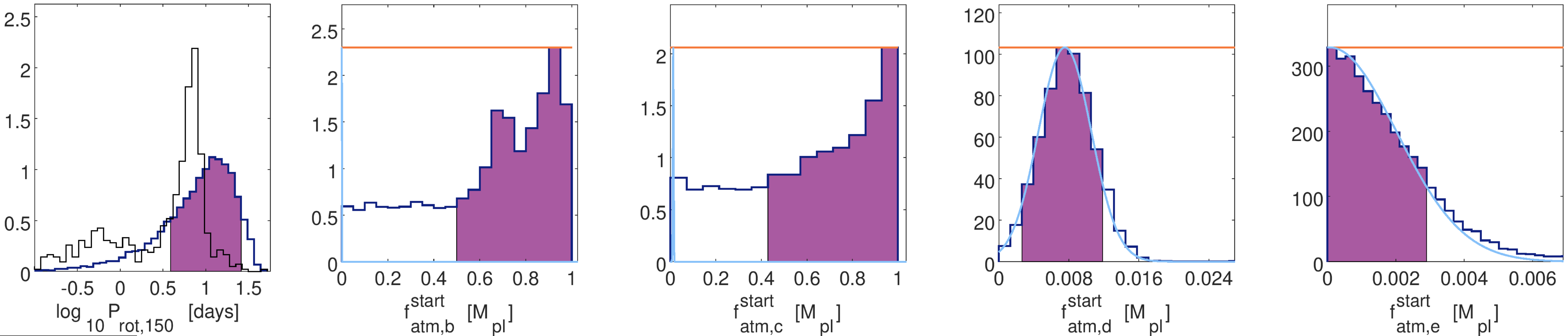}
  \caption{{\it From left to right:} Posterior distributions (dark blue lines) of the stellar rotation period at an age of $150$~Myr and of the initial atmospheric mass fractions of TOI-561 b, c, d and e. 
  In each panel, the purple region represents the $68$ per cent highest probability density intervals.
  In the left panel, the black thin line shows the rotation period distribution of stars member of open clusters with ages around $150$~Myr. Data are taken from \citet{johnstone15Prot150}, who report the rotation period of $\sim \, 2000$ stars belonging to the Pleiades, M50, M35, and NGC~2516, whose ages are between $125$ and $150$ Myr. To generate the black histogram we selected a sub-sample of $578$ stars, which have masses that deviate from \mstar\ less than $0.1$~\msun. In the other panels, the horizontal orange lines mark the uniform prior used in the fit, scaled to the highest peak of each posterior distribution for better visualization. The light blue lines indicate the current atmospheric mass fraction of each planet determined as described in Section~\ref{sec:internal_model}.
    }
    \label{fig:atmosphere}
\end{figure*}

Figure~\ref{fig:atmosphere} shows the results of the planetary atmospheric evolution simulations. As a proxy for the evolution of the stellar rotation period, in Figure~\ref{fig:atmosphere}, we show the posterior distribution of the stellar rotation period at an age of 150\,Myr, further comparing it to the distribution of stellar rotation periods observed in stars member of young clusters of comparable age and with masses that deviate from \mstar\ less than $0.1$~\msun\ \citep[from][]{johnstone15Prot150}. 
The inferred posterior distribution for the rotation period is consistent with membership of the slowly-rotating period-colours sequence in clusters of this age. 
However, this comparison should be taken with some caution, since there are no comprehensive studies on the rotation-colour distributions of $150$~Myr-old clusters with the same metallicity as TOI-561.
The initial atmospheric mass fractions of planets b and c are rather broad and peak at about one planetary mass. This is because both planets are close enough to the host star and have a small enough mass to have been subject to significant atmospheric escape. Therefore, to enable the presence of a thin hydrogen atmosphere, as predicted by the internal structure model, both planets had to host a significant hydrogen envelope after the formation and atmospheric accretion processes. Instead, planets d and e are far from the host star and massive enough not to have been subject to significant atmospheric escape, which is why we obtain an initial atmospheric mass fraction that resembles the current one. We also find that the posterior distributions of all input parameters match well the inserted priors (not shown here). As a whole, the results indicate that the currently observed system parameters are compatible with a scenario in which migration happened (if at all) exclusively inside the protoplanetary disk. Otherwise the code would have led to mismatches between the prior and posterior of the input parameters (particularly for what concerns the planetary masses and/or the stellar mass and age), in addition to showing incoherent results in the posterior distribution of the output parameters. This is for example the case of the TOI-1064 system, which is composed by two transiting planets with comparable masses and irradiation levels, but significantly different radii \citep{wilson2022}. In our framework in which planets do not migrate after the dispersal of the protoplanetary nebula, reproducing the physical parameters of the planets composing the TOI-1064 system requires different evolutions of the stellar rotation rate, which is not possible, thus calling for a post-nebula migration.

\section{Discussion and conclusions}
\begin{figure}
\centering
  \includegraphics[width=\linewidth]{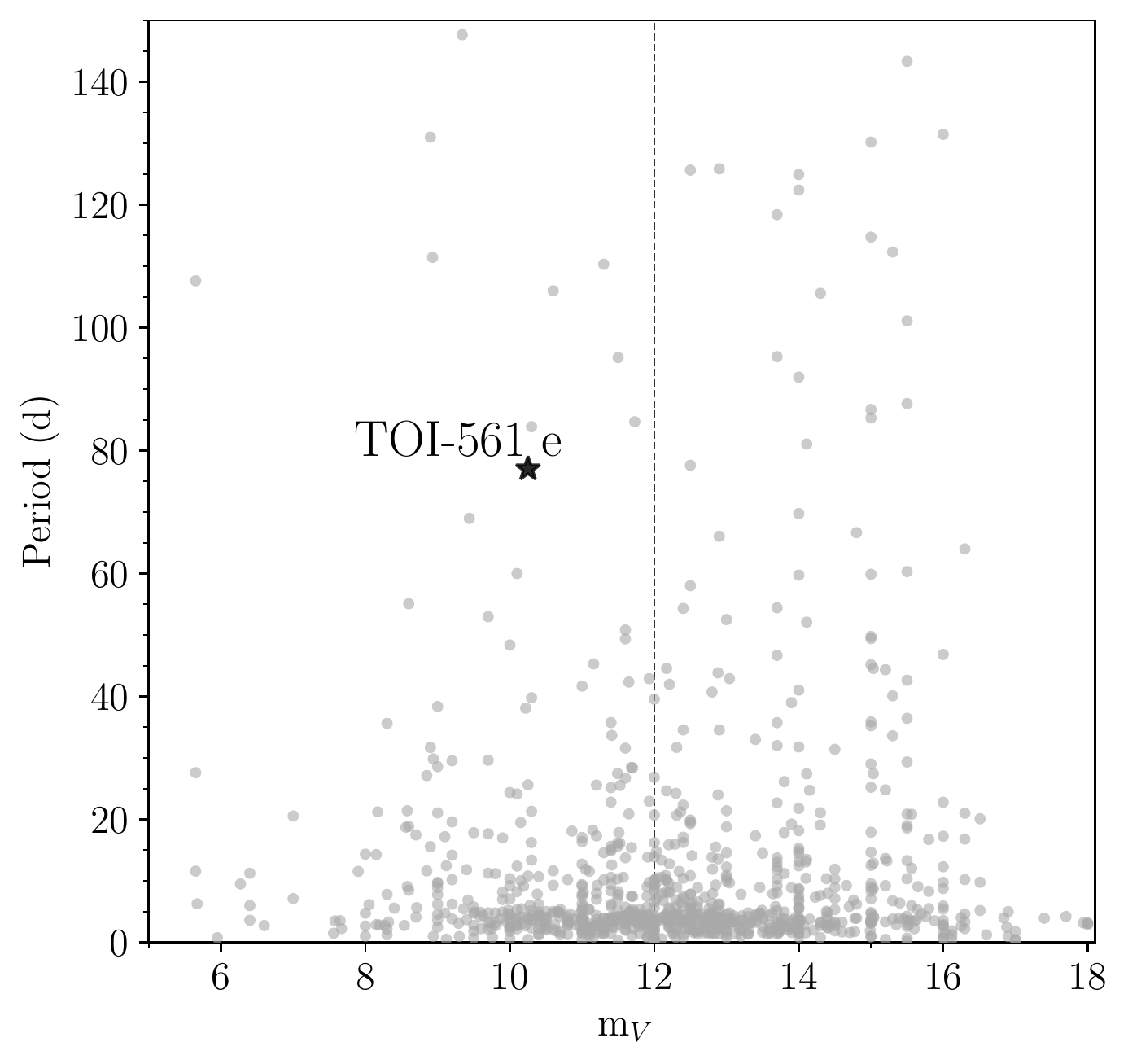}
  \caption{$V$ magnitude versus planetary periods for confirmed transiting exoplanets as reported in the Extrasolar Planets Encyclopaedia catalogue (\url{http://exoplanet.eu/catalog/}) in date 18 October 2021. The dashed vertical line marks $V=12$~mag. TOI-561 e is one of the few long-period planets orbiting a star bright enough for precise RV characterization.
    }
    \label{fig:Vmag_P}
\end{figure}

{\label{sec:conclusion}}
In this study, we confirm the presence of four transiting planets around TOI-561, with orbital periods of approximately $0.44$, $10.8$, $25.7$, and $77$~days (Table~\ref{table:joint_parameters}). 
Our analysis disproves the presence of the previously suggested planet TOI-561 f ($P \sim 16.3$~day; W21). 
TOI-561 is one of the few 4-planet systems having precise radius and mass measurements for all the planets. 
Thanks to our global photometric and RV analysis, we refined all masses and radii with respect to the L21 values, and we precisely determined the planetary bulk densities, with uncertainties of $14.4$\%, $13.6$\%, $10.2$\%, and $18.4$\% for planets b, c, d, and e, respectively. 
The higher uncertainty on planet e reflects the lower precision in the radius determination ($5$\% uncertainty), which is based on the analysis of a single {\it TESS} transit, and highlights the importance of the high-precision {\it CHEOPS} photometry. 
In fact, with a single {\it CHEOPS} transit we managed to decrease the uncertainty on the radius of planet d from $5.1$\% (L21, based on one {\it TESS} transit) to $2.5$\%. Including also the improvement on the mass, this implied a decrease on the density uncertainty from $18.9$\% to $10.2$\%.
We expect a similar improvement for planet e with future {\it CHEOPS} observations scheduled for 2022. 
The improvement in the radius of TOI-561 e is particularly important, since the planet is an interesting target for the study of the internal structure of cold sub-Neptunes.
Its long period ($P_{\rm d} = 77.03_{-0.24}^{+0.25}$~d) implies an insolation flux of $S_{\rm e} = 4.96 \pm 0.28$~$S_{\oplus}$ and a relatively cool zero Bond albedo equilibrium temperature of $T_{\rm eq, e} = 415 \pm 6$~K. As shown in Figure~\ref{fig:Vmag_P}, TOI-561 e is one of the few cool, long-period planets orbiting a star bright enough for precise RV characterization, and it is therefore an optimal test-case to refine tools and models that will be useful to characterize targets of future long-staring missions like {\it PLATO}.

TOI-561 hosts one of the most intriguing USP planets discovered to date.
As initially suggested by L21, our analysis confirms that TOI-561 b is the lowest density ($\rho_{\mathrm b} = 3.8 \pm 0.5$~\gcm) USP super-Earth that we know of (see Figure~\ref{fig:USPs_MR}), and it paves the way for in-depth studies of interior composition, and formation and evolution processes of USP planets. 
Even though now the mass values are consistent within $1\sigma$, contrary to what proposed by W21 (see Section~\ref{sec:pl_system}) TOI-561 b is not consistent with a pure rocky composition, and to explain the planetary density our internal structure modelling (Section~\ref{sec:internal_model}) predicts basically no H/He envelope, and a massive water layer. 
In this regard, an important point to consider is that, with an insolation flux of $S_{\rm b} \simeq 4745$~$S_{\oplus}$, the planet receives more irradiation from the star than the theoretical runaway greenhouse limit \citep{kasting1993, Goldblatt2012, kopparapu2013}.
In this case, a large water content would imply the presence of an extended steam atmosphere, which in turn would increase the measured radius with respect to a purely condensed water world, leading in our model to an overestimation of the bulk water content \citep{turbet2020}. 
The presence of a water steam envelope could eventually be tested with the {\it James Webb Space Telescope} ({\it JWST}). In fact, with an Emission Spectroscopy Metric (ESM, \citealt{kempton2018}) value of $8.2$, TOI-561 b is a promising target for secondary eclipse and phase curve observations.
More complex models, including a lighter core compositions (i.e. a Ca/Al enriched core), the modelling of water steam envelopes, or wet-melt solid interiors related to deep water reservoirs \citep{Dorn2021}, could be an interesting step forward in the understanding of the planet structure and composition. 
The low density of TOI-561 b could also be related to the fact that the host star is a metal-poor, thick-disk star.
\cite{adibekyan2021} showed that the composition of the rocky planets reflects the chemical abundances of the host star (even though not in a one-to-one relation), so implying a lighter composition for TOI-561 b with respect to other USP planets that orbit more metal-rich stars\footnote{All the USP planets shown in Figure~\ref{fig:MR_diagram} have \gfeh$> -0.14$.}. According to \citet{adibekyan2021}, the low density of TOI-561 b is consistent with the general $\rho$/$\rho_{\rm Earth-like}$~--~$f_{\rm iron}^{\rm star}$ trend and dispersion inferred from the sample of rocky planets analysed by the authors (see Figs. 2, 3 therein), where $\rho$/$\rho_{\rm Earth-like}$ is the planetary density normalised to that expected for an Earth-like composition, and $f_{\rm iron}^{\rm star}$ is the iron-to-silicate mass fraction of the protoplanetary disk as inferred from the stellar properties. An additional interesting remark concerns the Galactic kinematics of the host star. According to our analysis, performed as described in \citet{mustill2021}, TOI-561 is located in a low-density region of the 6-dimensional Galactic phase space (see \citealt{winter2020}, \citealt{mustill2021}, and \citealt{Kruijissen2021} for definition and discussion), 
which is not surprising given that TOI-561 is a thick disk star \citep{mustill2021}.
\citet{Kruijssen2020} showed that stars in low-density regions seem to host no super-Earths, but only sub-Neptunes, i.e. planets having a significant H/He envelope and therefore located above the radius gap. 
In this context, TOI-561 b is an interesting object that runs counter to this finding.
We point out that this result should be taken with some caution, since the \citet{Kruijssen2020} sample does not include planets with periods shorter than one day, and it excludes stars with ages $> 4.5$~Gyr\footnote{We note however that the stellar ages used in \citet{Kruijssen2020} are quite inhomogeneous, coming directly from the NASA Exoplanet Archive, and can therefore show a large scatter with respect to a homogeneous determination \citep{adibekyan2021_b}.}.

All the four planets seem to host a large water layer (Section~\ref{sec:internal_model}), although with high uncertainties, especially for planet c and d, due to the degeneracy related to the possible presence of a gas envelope. 
Also in this case, the presence of a considerable amount of water could be linked with the stellar properties.
In fact, \citet{santos2017} showed that metal-poor, thick disk stars are expected to form planetary building blocks with a higher water mass fraction ($\sim 76$\%) compared to metal-rich, thin disk stars ($\sim 58$\%).
Therefore, we would expect these stars to produce water-rich planets, a result that is in agreement with our findings on the TOI-561 system.

Except for TOI-561 b, all the other planets are suggested to host a non-negligible H/He envelope. In particular, the gas content of planet c ($\sim 1.3$ wt\%, the highest mass fraction among the four planets) implies a much lower density with respect to the density of planet d, even though the two planets have a similar size.
This is reflected in the different positions of the planets in the mass-radius diagram (Figure~\ref{fig:MR_diagram}). 
The two planets show hints of a different evolution for what concerns their gas content. 
In fact, our atmospheric evolution model (Section~\ref{sec:atmo_model}) suggests that planet c underwent a strong envelope loss after the atmospheric accretion and the dispersal of the protoplanetary nebula, while planet d (as well as planet e) did not experience strong atmospheric escape, with a current gas content that is comparable to the original one. 
The surprising difference in gas mass fraction between planets c, d and e, not only at present time but also at the end of their formation phase, takes probably its origin in the conditions that prevailed during the protoplanetary disk phase. Planet c is indeed likely sub-critical because of its low mass, where sub-critical planets are those with masses below the critical value required to initiate runaway gas accretion (see \citealt{helled2014} for a recent review on the core accretion model), whereas planets d and e never accreted large amounts of gas as demonstrated in Section~\ref{sec:atmo_model}, and so they also remained always below the critical mass. The interpretation of the different gas mass fractions could therefore result from the structure of sub-critical planets. In this case, the gas mass fraction depends on the core mass, the thermodynamical properties in the disk, and more importantly the accretion rate of solids (lower accretion rate translating in larger gas mass fraction). Interpreting the internal structure of the four planets of the system in a global planetary system formation model could therefore constrain these parameters.

With its derived properties, TOI-561 c has a Transmission Spectroscopy Metric (TSM, \citealt{kempton2018}) of $110.4$, and is therefore a suitable target for atmospheric characterization with {\it JWST}.\footnote{\cite{kempton2018} suggest to select planets with TSM $>92$ for $1.5$~\rearth $<$ \rplanet~$<2.75$~\rearth, and TSM $>84$ for $2.75$~\rearth $<$ \rplanet~$<4$~\rearth.}
Instead, planets d and e have lower TSM values of $30.7$ and $16.2$, respectively. 
As the TSM is proportional to the equilibrium temperature, it is not surprising to obtain lower values for the two planets, given their longer periods.

In addition to the characterization of the four planets, we also identified a significant long-term signal ($P \sim 473$~d) in the RVs. On the basis of our current dataset, we are not able to distinguish between a stellar (magnetically-induced) or planetary origin.
Long-term monitoring using both spectroscopic ground-based facilities and future long-staring missions like the {\it PLATO} spacecraft will allow us to shed light on the nature of this additional signal, and to potentially find new outer companions.
It is worth noting that, if the above-mentioned signal proves in future to be of planetary origin, there is a non-zero chance that, under the assumption of co-planarity, such a planet would transit. In fact, assuming the same inclination of planet e and using the semi-major axis $a/\rstar = 279_{-10}^{+14}$ derived from our global fit, we infer an impact parameter of $0.97_{-0.63}^{+0.49}$. Moreover, the planet would orbit in TOI-561's empirical habitable zone ($ 175 \lesssim P \lesssim 652$~d), as originally defined by \cite{kasting1993} using a 1D climate model, and later updated in \cite{kopparapu2013, ramirez2016} for main-sequence stars with $2600 <$~\teff~$<10000$~K.

This work bears witness to the fruitful results that can be obtained by the timely combination of data coming from different instruments. 
It adds to the works \citep{bonfanti2021, leleu2021, delrez2021} that prove the potential of {\it CHEOPS} in precisely characterizing {\it TESS}-discovered exoplanets, as well as demonstrating the key role of high-precision spectrographs such as HARPS-N when working in synergy with space-based facilities.

\section*{Acknowledgements}
We thank the referee for the useful comments that helped improving the quality of the manuscript.
We gratefully acknowledge the MuSCAT2 team for the availability in collaborating with the \cheops\ Consortium on target monitoring, and the NGTS team for providing ground-based photometry that helped the scheduling of our observations.
\cheops\ is an ESA mission in partnership with Switzerland with
important contributions to the payload and the ground segment
from Austria, Belgium, France, Germany, Hungary, Italy, Portugal,
Spain, Sweden, and the United Kingdom. 
The \cheops\ Consortium gratefully acknowledge the support received by
all the agencies, offices, universities, and industries involved.
Their flexibility and willingness to explore new approaches were
essential to the success of the mission. 
This work is based on observations made with the Italian 
Telescopio Nazionale Galileo (TNG) operated on the island of La Palma
by the Fundaci\'on Galileo Galilei of the INAF
at the Spanish Observatorio del Roque de los
Muchachos of the Instituto de Astrofisica de Canarias
(GTO program, and A40TAC\_23 program from INAF-TAC).
The HARPS-N project was funded by the Prodex Program of
the Swiss Space Office (SSO), the Harvard University Origin
of Life Initiative (HUOLI), the Scottish Universities Physics
Alliance (SUPA), the University of Geneva, the Smithsonian
Astrophysical Observatory (SAO), and the Italian National
Astrophysical Institute (INAF), University of St. Andrews,
Queen's University Belfast and University of Edinburgh.
This paper includes data collected by the {\it TESS} mission,
which are publicly available from the Mikulski Archive for Space
Telescopes (MAST). 
Funding for the {\it TESS} mission is provided 
by the NASA Explorer Program. 
Resources supporting this work were provided by the NASA High-End Computing (HEC) Program through the NASA Advanced Supercomputing (NAS) Division at Ames Research Center for the production of the SPOC data products.
This research has
made use of the NASA Exoplanet Archive, which is
operated by the California Institute of Technology, 
under contract with the National Aeronautics and Space
Administration under the Exoplanet Exploration Program. 
This research has made use of data obtained from the portal \url{http://www.exoplanet.eu/} of The Extrasolar Planets Encyclopaedia.
This work has made use of data from the European Space Agency (ESA) mission
{\it Gaia} (\url{https://www.cosmos.esa.int/gaia}), processed by the {\it Gaia}
Data Processing and Analysis Consortium (DPAC,
\url{https://www.cosmos.esa.int/web/gaia/dpac/consortium}). 
Funding for the DPAC
has been provided by national institutions, in particular the institutions
participating in the {\it Gaia} Multilateral Agreement.
This publication makes use of data products from the Two
Micron All Sky Survey, which is a joint project of the
University of Massachusetts and the Infrared Processing and
Analysis Center/California Institute of Technology, funded by
the National Aeronautics and Space Administration and the
National Science Foundation. 
GL acknowledges support by CARIPARO Foundation, according to the agreement CARIPARO-Universit{\`a} degli Studi di Padova (Pratica n. 2018/0098).
TW and ACC acknowledge support from STFC consolidated grant numbers ST/R000824/1 and ST/V000861/1, and UKSA grant number ST/R003203/1.
YA, MJH, B.-O.D. and M.L. acknowledge the support of the Swiss National Fundation under grants 200020\_172746, PP00P2-190080, and PCEFP2\_194576.
SH gratefully acknowledges CNES funding through the grant 837319. 
GPi, VNa, GSs, IPa, LBo, and RRa acknowledge the funding support from Italian Space Agency (ASI)
regulated by ‘‘Accordo ASI-INAF n. 2013-016-R.0 del 9 luglio 2013 e integrazione del 9 luglio 2015 \cheops\ Fasi A/B/C’’.
ADe acknowledges support from the European Research Council (ERC) under the European Union's Horizon 2020 research and innovation programme (project {\sc Four Aces}, grant agreement No. 724427), and from the National Centre for Competence in Research ‘‘PlanetS’’ supported by the Swiss National Science Foundation (SNSF).
KR is grateful for support from the UK STFC via grant ST/V000594/1.
This work has been supported by the National Aeronautics and Space Administration under grant No. NNX17AB59G, issued through the Exoplanets Research Program.
S.S. has received funding from the European
Research Council (ERC) under the European Union’s Horizon 2020 research
and innovation programme (grant agreement No 833925, project STAREX).
M.G. is an F.R.S.-FNRS Senior Research Associate. V.V.G. is an F.R.S-FNRS Research Associate. 
L.D. is an F.R.S.-FNRS Postdoctoral Researcher.
This work has been carried out within the framework of the NCCR PlanetS supported by the Swiss National Science Foundation.
AMu and MF acknowledge support from the Swedish National Space Agency (career grant 120/19C, DNR 65/19, 174/18).
ABr was supported by the SNSA. 
We acknowledge support from the Spanish Ministry of Science and Innovation and the European Regional Development Fund through grants ESP2016-80435-C2-1-R, ESP2016-80435-C2-2-R, PGC2018-098153-B-C33, PGC2018-098153-B-C31, ESP2017-87676-C5-1-R, MDM-2017-0737 Unidad de Excelencia Maria de Maeztu-Centro de Astrobiología (INTA-CSIC), as well as the support of the Generalitat de Catalunya/CERCA programme. 
The MOC activities have been supported by the ESA contract No. 4000124370. 
S.G.S., S.C.C.B. and V.A. acknowledge support from FCT through FCT contract nr. CEECIND/00826/2018, POPH/FSE (EC), nr. IF/01312/2014/CP1215/CT0004, and IF/00650/2015/CP1273/CT0001, respectively.
O.D.S.D. is supported in the form of work contract (DL 57/2016/CP1364/CT0004) funded by national funds through FCT. 
XB, SC, DG, MF and JL acknowledge their role as ESA-appointed \cheops\ science team members. 
The Belgian participation to \cheops\ has been supported by the Belgian Federal Science Policy Office (BELSPO) in the framework of the PRODEX Program, and by the University of Liège through an ARC grant for Concerted Research Actions financed by the Wallonia-Brussels Federation. 
This work was supported by FCT~--~Fundação para a Ciência e a Tecnologia through national funds and by
FEDER through COMPETE2020~--~Programa Operacional Competitividade e Internacionalizacão by these grants: UID/FIS/04434/2019, UIDB/04434/2020, UIDP/04434/2020, PTDC/FIS-AST/32113/2017 \& POCI-01-0145-FEDER-032113, PTDC/FIS-AST/28953/2017 \& POCI-01-0145-FEDER-028953, PTDC/FIS-AST/28987/2017 \& POCI-01-0145-FEDER-028987.
This project has received funding from the European Research Council (ERC) under the European Union’s Horizon 2020 research and innovation programme (project {\sc Four Aces}, grant agreement No 724427).
DG and LMS gratefully acknowledge financial support from the CRT foundation under Grant No. 2018.2323 ``Gaseousor rocky? Unveiling the nature of small worlds''. 
KGI is the ESA \cheops\ Project Scientist and is responsible for the ESA \cheops\ Guest Observers Programme. She does not participate in, or contribute to, the definition of the Guaranteed Time Programme of the \cheops\ mission through which observations described in this paper have been taken, nor to any aspect of target selection for the programme. 
This work was granted access to the HPC resources of MesoPSL financed by the Region Ile de France and the project Equip@Meso (reference ANR-10-EQPX-29-01) of the programme Investissements d'Avenir supervised by the Agence Nationale pour la Recherche. 
PM acknowledges support from STFC research grant number ST/M001040/1. 
This work was also partially supported by a grant from the Simons Foundation (PI Queloz, grant number 327127).
GyMSz acknowledges the support of the Hungarian National Research, Development and Innovation Office (NKFIH) grant K-125015, a PRODEX Institute Agreement between the ELTE E\"otv\"os Lor\'and University and the European Space Agency (ESA-D/SCI-LE-2021-0025), the Lend\"ulet LP2018-7/2021 grant of the Hungarian Academy of Science and the support of the city of Szombathely.
This work is partly supported by JSPS KAKENHI Grant Number JP18H05439, JST CREST Grant Number JPMJCR1761, the Astrobiology Center of National Institutes of Natural Sciences (NINS) (Grant Number AB031010).
E. E-B. acknowledges financial support from the European Union and the State Agency of Investigation of the Spanish Ministry of Science and Innovation (MICINN) under the grant PRE2020-093107 of the Pre-Doc Program for the Training of Doctors (FPI-SO) through FSE funds.

\section*{Data Availability}
HARPS-N observations and data products are available through the Data \& Analysis Center for Exoplanets (DACE) at \url{https://dace.unige.ch/}. {\it TESS} data products can be accessed through the official NASA website \url{https://heasarc.gsfc.nasa.gov/docs/tess/data-access.html}. 
All underlying data are available either in the appendix/online supporting material or will be available via VizieR at CDS.


\bibliographystyle{mnras}
\bibliography{Bibliography} 




\appendix

\section{CHEOPS light curves and telegraphic pixel treatment}\label{sec:app_tel_pix}
\begin{figure}
\centering
  \includegraphics[width=\linewidth]{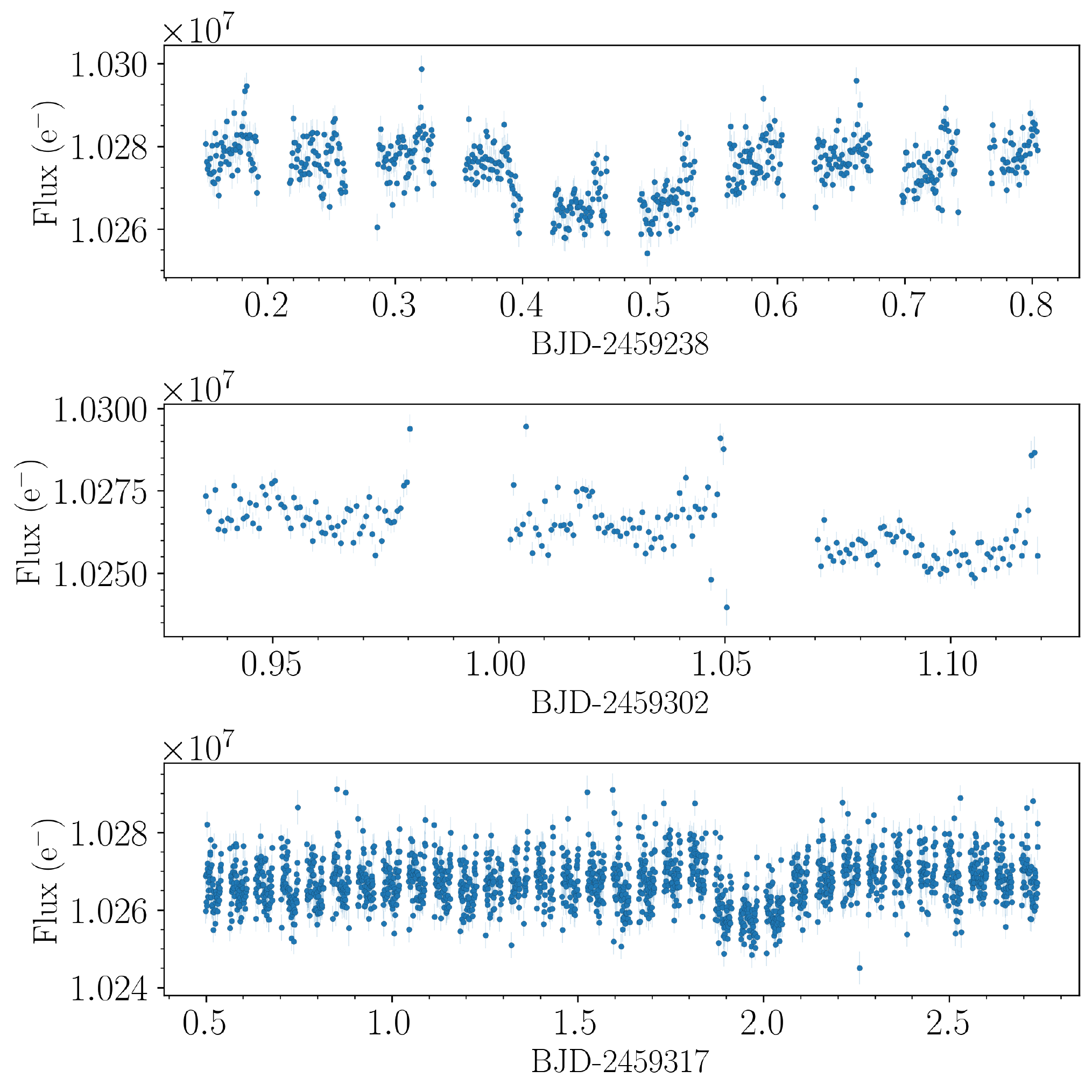}
  \caption{{\it CHEOPS} RINF light curves of TOI-561 as extracted from the DRP, with $4 \sigma$-clipping for outliers removal. Visits $1$, $2$ and $3$ are shown from top to bottom.
    }
    \label{fig:cheops_raw}
\end{figure}

As described in Section~\ref{sec:cheops_observations}, the three {\it CHEOPS} visits of TOI-561 were reduced via the standard DRP processing. The light curves presented in this study, obtained using the RINF aperture size (RINF $= 0.9 \, \times$ DEFAULT, where DEFAULT = $25$~px; see also Section~\ref{sec:cheops_observations}), are shown in Figure~\ref{fig:cheops_raw}. 
While for the two initial visits the automatic DRP processes was performed, the appearance of some telegraphic pixels during the third visit required a more in-depth analysis.

\begin{figure}
\centering
  \includegraphics[width=\linewidth]{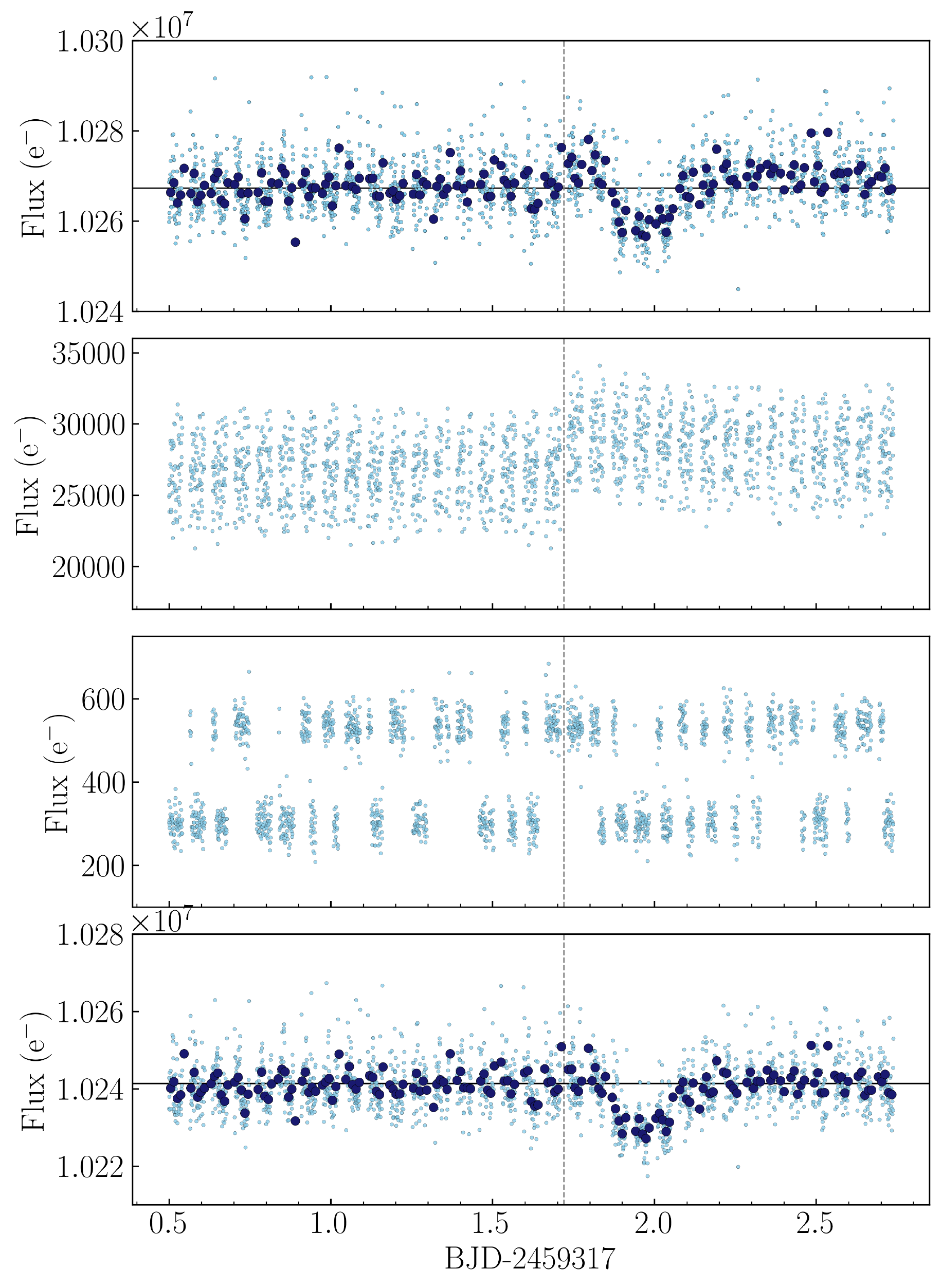}
  \caption{Top panel: TOI-561 RINF original light curve of the third visit (light blue dots) after the removal of $4\sigma$ outliers, with over-plotted the $15$-minute binned light curve (dark blue dots). The start of the flux jump due to the telegraphic pixel is marked with the dashed vertical line. 
  Second panel: light curve of the telegraphic pixel located within the {\it CHEOPS} PSF.
  Third panel: light curve of the two additional telegraphic pixels located within the RINF aperture.
  Bottom panel: corrected light curve after masking the three telegraphic pixels. 
    }
    \label{fig:telegraphic}
\end{figure}

In addition to the large number of known hot pixels present in the {\it CHEOPS} CCD (some of them visible in Figure~\ref{fig:cheops_fov}), some normal pixels can change their behaviour during the duration of a visit, for example becoming ‘hot’ after a SAA crossing.
These pixels, called ‘telegraphic’ for their abnormal behaviour, can affect the photometry if located within the photometric aperture (see for example \citealt{leleu2021}). 
During the third {\it CHEOPS} visit, we identified an unusual flux bump before the ingress of TOI-561 d transit, at BJD~$\sim 2459318.75$ (top panel, Figure~\ref{fig:telegraphic}). 
After analyzing the statistics of each pixel light curve within the photometric aperture, we detected a telegraphic pixel with a large flux variation (second panel, Figure~\ref{fig:telegraphic}) located within the {\it CHEOPS} PSF. The exact position of this pixel on the \cheops\ CCD is shown in Figure~\ref{fig:cheops_fov}. 
We masked the pixel flux and repeated the photometric extraction of the visit using the RINF aperture, so removing the flux jump in the light curve (bottom panel, Figure~\ref{fig:telegraphic}). 
During this analysis, we detected two additional telegraphic pixels within the photometric aperture, inducing smaller, but still significant variations in the light curve flux (third panel, Figure~\ref{fig:telegraphic}).
We corrected for the effect of these pixels as described above.

\begin{figure}
\centering
  \includegraphics[width=\linewidth]{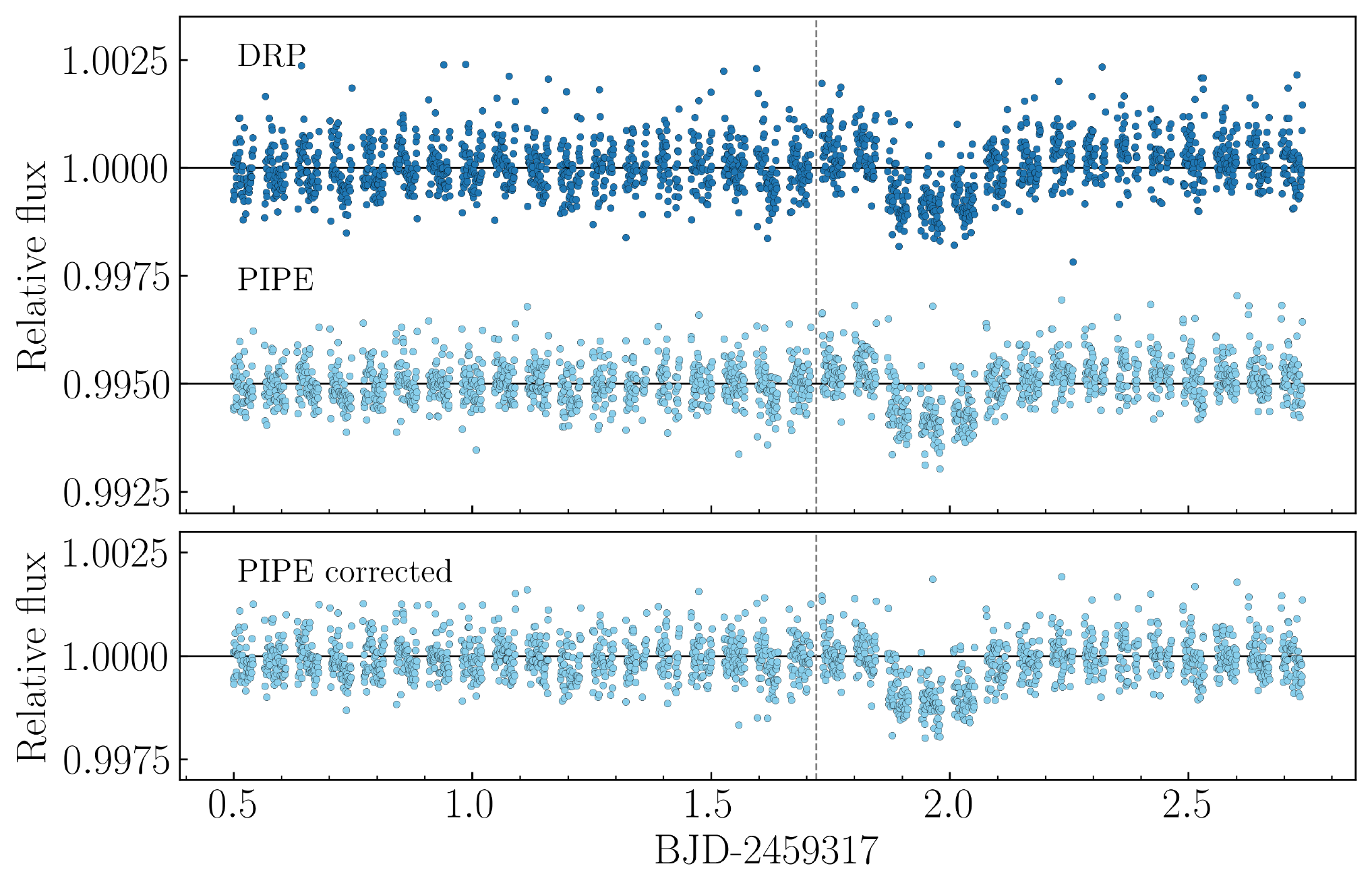}
  \caption{Top panel: comparison between DRP and \texttt{PIPE}-extracted light curve of TOI-561 third visit, before the telegraphic pixel correction and with $4\sigma$ outliers removal. The DRP has an MAD of $371$ ppm over the whole visit, while \texttt{PIPE} of $325$ ppm. Bottom panel: \texttt{PIPE} light curve after the telegraphic pixel correction. The light curve gets slightly noisier (MAD~$=331$~ppm) because one less pixel is considered in the reduction, but more reliable thanks to the exclusion of the telegraphic pixel flux. In both panels, the beginning of the flux jump is highlighted with a vertical dashed line.
    }
    \label{fig:cheops_PIPE}
\end{figure}

While investigating the nature of the flux bump happening during the third visit, we also extracted the light curve using a PSF-photometry approach exploiting the \texttt{PIPE} (PSF Imagette Photometric Extraction) software\footnote{\url{https://pipe-cheops.readthedocs.io/en/latest/index.html}}. 
\texttt{PIPE} is a photometric extraction package specifically developed to extract {\it CHEOPS} light curves by applying PSF photometry on the $60$-pixel imagettes, complementing the offical DRP extraction. 
The use of PSF photometry makes usually easier to filter out the impact of hot pixels and cosmic rays, by either giving them a lower weight or masking them entirely in the fitting process.
However, in this case the telegraphic pixel was located inside the {\it CHEOPS} PSF, requiring a careful manual masking.
As for the DRP light curve,  the flux bump in the \texttt{PIPE} photometry is reduced after masking the telegraphic pixel (bottom panel, Figure~\ref{fig:cheops_PIPE}).
The \texttt{PIPE}-extracted light curve resulted in a slightly lower mean absolute deviation (MAD) with respect to the DRP photometry (top panel, Figure~\ref{fig:cheops_PIPE}), mainly due to the lower number of  outliers present in the PSF photometry. For a more detailed comparison between \texttt{PIPE} and DRP photometries, see \cite{morris2021}. We performed the same global analysis described in Section~\ref{sec:global_analysis} using the \texttt{PIPE} light curve instead of the DRP one, obtaining consistent results and comparable uncertainties on the transit parameters of both planets b and d. 
We therefore decided to use the light curve obtained with the official DRP extraction in our final analysis. 

\section{Floating Chunk Offset method on TOI-561 b}\label{sec:app_FCO}

In order to investigate the literature discrepancy on the mass of TOI-561 b (Section~\ref{sec:pl_system}), we adopted a specific observing strategy with HARPS-N targeting the USP planet (Section~\ref{sec:harpn_observations}), obtaining multiple observations during the same night for $22$ nights. 
Multiple nightly observations can be used to precisely infer the mass of USP planets using the Floating Chunk Offset method (FCO; \citealt{Hatzes2014}), which consists in applying a nightly offset to remove all the other signals present in the system, both of planetary and stellar origin (i.e. \citealt{Howard2013, Pepe2013, Malavolta2018, Frustagli2020}).
The FCO method is only applicable when the separation between the USP period and the period of all the other signals is large enough, and the RV semi-amplitude has a similar or larger value with respect to the other signals. 
As demonstrated in L21, these conditions apply to TOI-561 b, for which the authors derived an FCO semi-amplitude of $K_{\rm b, FCO} = 1.80 \pm 0.38$~\ms\ ($M_{\rm b, FCO} = 1.83 \pm 0.39$~\mearth) exploiting multiple observations collected over ten nights.

\begin{figure}
\centering
  \includegraphics[width=0.9\linewidth]{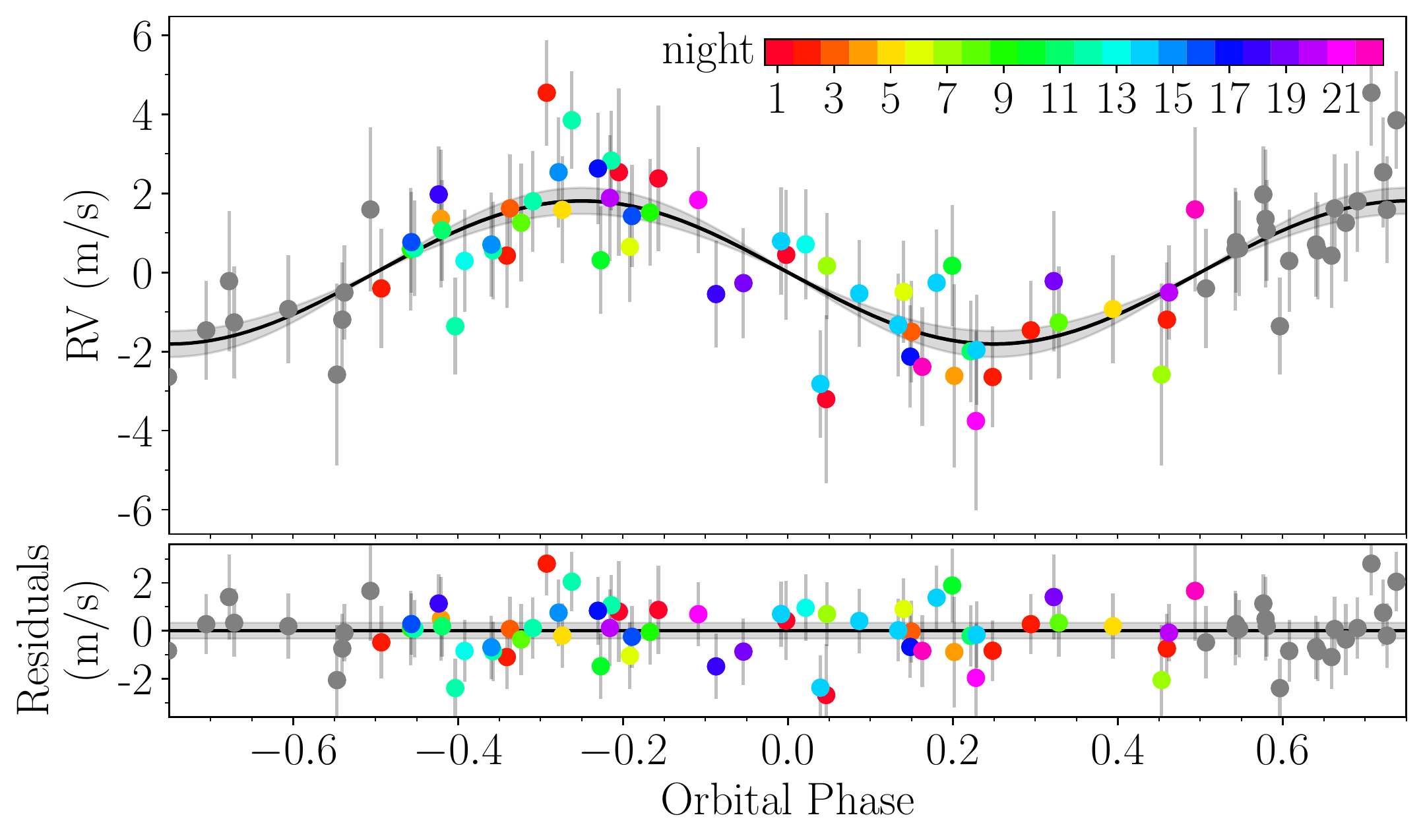}
  \caption{Phase-folded RVs of the $22$ HARPS-N nights used to model TOI-561 b with the FCO method. The error bars include the jitter term added in quadrature. 
    }
    \label{fig:USP}
\end{figure}

Here, we applied the FCO method to TOI-561 b on a total of $22$ HARPS-N nights, adding $12$ novel nights to the $10$ nights already presented in L21. Out of the total set, four nights have six multiple observations extending over more than $40$ per cent of the orbital period of the planet, and span opposite orbital phases to provide an optimal phase coverage.
We performed a \texttt{PyDE} + \texttt{emcee} fit with \texttt{PyORBIT}, assuming a fixed zero eccentricity and Gaussian priors on period and $T_0$ coming from the global fit, and we added a jitter term to account for possible additional white noise. 
We derived a semi-amplitude of $K_{\rm b} = 1.81 \pm 0.31$~\ms, corresponding to a mass of $M_{\rm b} = 1.86 \pm 0.33$~\mearth, with a jitter of $0.96_{-0.23}^{+0.25}$~\ms. 
Figure~\ref{fig:USP} shows the resulting phase-folded RVs. 
The derived mass and semi-amplitude are nicely in agreement with the L21 values, and they support the values inferred from our joint photometric and RV modelling (Section~\ref{sec:global_analysis}), being consistent within $1\sigma$. 
Given the higher number of RVs included in the joint fit, which led to smaller uncertainties on the derived parameters, we decided to adopt as final values for TOI-561 b the ones obtained from the global modelling, i.e. $K_{\rm b} = 1.93 \pm 0.21$~\ms, $M_{\rm b} = 1.99 \pm 0.22$~\mearth.

\section*{Affiliations}
\noindent
{\it
$^{1}$ Dipartimento di Fisica e Astronomia "Galileo Galilei", Università degli Studi di Padova, Vicolo dell'Osservatorio 3, 35122 Padova, Italy\\
$^{2}$ INAF, Osservatorio Astronomico di Padova, Vicolo dell'Osservatorio 5, 35122 Padova, Italy\\
$^{3}$ Centre for Exoplanet Science, SUPA School of Physics and Astronomy, University of St Andrews, North Haugh, St Andrews KY16 9SS, UK\\
$^{4}$ Physikalisches Institut, University of Bern, Gesellsschaftstrasse 6, 3012 Bern, Switzerland\\
$^{5}$ Center for Space and Habitability, Gesellsschaftstrasse 6, 3012 Bern, Switzerland\\
$^{6}$ Astrophysics Group, Cavendish Laboratory, University of Cambridge, J.J. Thomson Avenue, Cambridge CB3 0HE, UK\\
$^{7}$ Kavli Institute for Cosmology, University of Cambridge, Madingley Road, Cambridge CB3 0HA, UK\\
$^{8}$ Space Research Institute, Austrian Academy of Sciences, Schmiedlstrasse 6, A-8042 Graz, Austria\\
$^{9}$ Astrophysics Group, University of Exeter, Exeter EX4 2QL, UK\\
$^{10}$ Aix Marseille Univ, CNRS, CNES, LAM, 38 rue Frédéric Joliot-Curie, 13388 Marseille, France\\
$^{11}$ Department of Astronomy, University of Geneva, Chemin Pegasi 51, Versoix, Switzerland\\
$^{12}$ Department of Physics and Kavli Institute for Astrophysics and Space Research, Massachusetts Institute of Technology, Cambridge, MA 02139, USA\\
$^{13}$ Department of Astronomy, University of Wisconsin-Madison, Madison, WI 53706, USA\\
$^{14}$ Center for Astrophysics | Harvard and Smithsonian, 60 Garden Street, Cambridge, MA 02138, USA\\
$^{15}$ SUPA, Institute for Astronomy, University of Edinburgh, Blackford Hill, Edinburgh EH9 3HJ, Scotland, UK\\
$^{16}$ Centre for Exoplanet Science, University of Edinburgh, Edinburgh EH93FD, UK\\
$^{17}$ Department of Astronomy, Stockholm University, AlbaNova University Center, 10691 Stockholm, Sweden\\
$^{18}$ Fundación Galileo Galilei - INAF, Rambla J. A. F. Perez, 7, E-38712 S.C. Tenerife, Spain\\
$^{19}$ INAF – Osservatorio Astronomico di Brera, via E. Bianchi 46, I-23807 Merate (LC), Italy\\
$^{20}$ Instituto de Astrofisica e Ciencias do Espaco, Universidade do Porto, CAUP, Rua das Estrelas, 4150-762 Porto, Portugal\\
$^{21}$ INAF - Osservatorio Astrofisico di Torino, Via Osservatorio 20, I-10025 Pino Torinese, Italy\\
$^{22}$ Space sciences, Technologies and Astrophysics Research (STAR) Institute, Universit{\'e} de Liège, Allée du 6 Août 19C, 4000 Liège, Belgium\\
$^{23}$ University of Southern Queensland, Centre for Astrophysics, West Street, Toowoomba, QLD 4350 Australia\\
$^{24}$ Lund Observatory, Dept. of Astronomy and Theoretical Physics, Lund University, Box 43, 22100 Lund, Sweden\\
$^{25}$ Instituto de Astrofisica de Canarias, 38200 La Laguna, Tenerife, Spain\\
$^{26}$ Departamento de Astrofísica, Universidad de La Laguna, E-38206 La Laguna, Tenerife, Spain\\
$^{27}$ Institut de Ciencies de l'Espai (ICE, CSIC), Campus UAB, Can Magrans s/n, 08193 Bellaterra, Spain\\
$^{28}$ Institut d'Estudis Espacials de Catalunya (IEEC), 08034 Barcelona, Spain\\
$^{29}$ Admatis, 5. Kandó Kálmán Street, 3534 Miskolc, Hungary\\
$^{30}$ Depto. de Astrofisica, Centro de Astrobiologia (CSIC-INTA), ESAC campus, 28692 Villanueva de la Cañada (Madrid), Spain\\
$^{31}$ Departamento de Fisica e Astronomia, Faculdade de Ciencias, Universidade do Porto, Rua do Campo Alegre, 4169-007 Porto, Portugal\\
$^{32}$ Université Grenoble Alpes, CNRS, IPAG, 38000 Grenoble, France\\
$^{33}$ DTU Space, National Space Institute, Technical University of Denmark, Elektrovej 328, DK-2800 Kgs. Lyngby, Denmark\\
$^{34}$ Institute of Planetary Research, German Aerospace Center (DLR), Rutherfordstrasse 2, 12489 Berlin, Germany\\
$^{35}$ Université de Paris, Institut de physique du globe de Paris, CNRS, F-75005 Paris, France\\
$^{36}$ Centre for Mathematical Sciences, Lund University, 22100 Lund, Sweden\\
$^{37}$ Astrobiology Research Unit, Université de Liège, Allée du 6 Août 19C, B-4000 Liège, Belgium\\
$^{38}$ Department of Astronomy, Stockholm University, SE-106 91 Stockholm, Sweden\\
$^{39}$ Leiden Observatory, University of Leiden, PO Box 9513, 2300 RA Leiden, The Netherlands\\
$^{40}$ Department of Space, Earth and Environment, Chalmers University of Technology, Onsala Space Observatory, 43992 Onsala, Sweden\\
$^{41}$ Dipartimento di Fisica, Università degli Studi di Torino, via Pietro Giuria 1, I-10125, Torino, Italy\\
$^{42}$ Department of Astrophysics, University of Vienna, Tuerkenschanzstrasse 17, 1180 Vienna, Austria\\
$^{43}$ Division Technique INSU, CS20330, 83507 La Seyne sur Mer cedex, France\\
$^{44}$ Department of Physics, University of Warwick, Gibbet Hill Road, Coventry CV4 7AL, United Kingdom\\
$^{45}$ Science and Operations Department - Science Division (SCI-SC), Directorate of Science, European Space Agency (ESA), European Space Research and Technology Centre (ESTEC),
Keplerlaan 1, 2201-AZ Noordwijk, The Netherlands\\
$^{46}$ NASA Ames Research Center, Moffett Field, CA 94035, USA\\
$^{47}$ Konkoly Observatory, Research Centre for Astronomy and Earth Sciences, 1121 Budapest, Konkoly Thege Miklós út 15-17, Hungary\\
$^{48}$ ELTE E\"otv\"os Lor\'and University, Institute of Physics, P\'azm\'any P\'eter s\'et\'any 1/A, 1117 Budapest, Hungary\\
$^{49}$ IMCCE, UMR8028 CNRS, Observatoire de Paris, PSL Univ., Sorbonne Univ., 77 av. Denfert-Rochereau, 75014 Paris, France\\
$^{50}$ Institut d'astrophysique de Paris, UMR7095 CNRS, Université Pierre \& Marie Curie, 98bis blvd. Arago, 75014 Paris, France\\
$^{51}$ Astrophysics Group, Keele University, Staffordshire, ST5 5BG, United Kingdom\\
$^{52}$ INAF - Osservatorio Astronomico di Palermo, Piazza del Parlamento 1, I-90134 Palermo, Italy\\
$^{53}$ INAF - Osservatorio Astronomico di Cagliari, Via della Scienza 5, I-09047, Selargius, Italy\\
$^{54}$ Komaba Institute for Science, The University of Tokyo, 3-8-1 Komaba, Meguro, Tokyo 153-8902, Japan\\
$^{55}$ Astrobiology Center, 2-21-1 Osawa, Mitaka, Tokyo 181-8588, Japan\\
$^{56}$ INAF, Osservatorio Astrofisico di Catania, Via S. Sofia 78, 95123 Catania, Italy\\
$^{57}$ P\&P Software GmbH, High Tech Center, Taegerwilen, CH-8274, Switzerland\\
$^{58}$ Institute of Optical Sensor Systems, German Aerospace Center (DLR), Rutherfordstrasse 2, 12489 Berlin, Germany\\
$^{59}$ ESTEC, European Space Agency, 2201AZ, Noordwijk, NL\\
$^{60}$ Center for Astronomy and Astrophysics, Technical University Berlin, Hardenbergstrasse 36, 10623 Berlin, Germany\\
$^{61}$ Institut für Geologische Wissenschaften, Freie Universität Berlin, 12249 Berlin, Germany\\
$^{62}$ Department of Earth, Atmospheric, and Planetary Sciences, Massachusetts Institute of Technology, Cambridge, MA 02139, USA\\
$^{63}$ Department of Aeronautics and Astronautics, Massachusetts Institute of Technology, Cambridge, MA 02139, USA\\
$^{64}$ ELTE Eötvös Loránd University, Gothard Astrophysical Observatory, 9700 Szombathely, Szent Imre h. u. 112, Hungary\\
$^{65}$ MTA-ELTE Exoplanet Research Group, 9700 Szombathely, Szent Imre h. u. 112, Hungary\\
$^{66}$ SETI Institute, Mountain View, CA  94043, USA\\
$^{67}$ Institute of Astronomy, University of Cambridge, Madingley Road, Cambridge, CB3 0HA, United Kingdom\\
$^{68}$ Department of Astrophysical Sciences, Princeton University, 4 Ivy Lane, Princeton, NJ 08544, USA\\
}

\bsp	
\label{lastpage}
\end{document}